\documentclass[ms]{informs4}
\JOURNAL{Working Paper}
\JOURNALshort{}
\usepackage{eqndefns-left} 
\RequirePackage{tgtermes}
\RequirePackage{newtxtext}
\RequirePackage{newtxmath}
\RequirePackage{bm}
\RequirePackage{endnotes}

\OneAndAHalfSpacedXII 

\usepackage{soul}

\usepackage{algorithm}
\usepackage[normalem]{ulem}
\usepackage{multirow}
\usepackage{algpseudocode}
\usepackage{tikz}
\usepackage{soul}
\usepackage[hidelinks,breaklinks=true]{hyperref}  
\usepackage{xurl}  
\renewcommand{\url}[1]{\href{#1}{\nolinkurl{#1}}}

\usepackage{natbib}
 \bibpunct[, ]{(}{)}{,}{a}{}{,}%
 \def\bibfont{\footnotesize}%
\usepackage{xcolor}

\usepackage{subcaption}
\usepackage{graphicx}
\usepackage{booktabs}
\usepackage{array}
\newsavebox{\regtablebox}

\usetikzlibrary{positioning,shapes.geometric,arrows.meta}

\usepackage{enumitem}

\usepackage[framemethod=default]{mdframed}
\newmdenv[
  linewidth=0.5pt,
  linecolor=black!50,
  backgroundcolor=black!2,
  innerleftmargin=10pt,
  innerrightmargin=10pt,
  innertopmargin=8pt,
  innerbottommargin=8pt,
  skipabove=8pt,
  skipbelow=8pt,
  leftmargin=18pt,
  rightmargin=18pt
]{verbatimbox}

\newcolumntype{C}[1]{>{\centering\let\newline\\\arraybackslash\hspace{0pt}}m{#1}}
\newcolumntype{L}[1]{>{\raggedright\let\newline\\\arraybackslash\hspace{0pt}}m{#1}}
\newcolumntype{R}[1]{>{\raggedleft\let\newline\\\arraybackslash\hspace{0pt}}m{#1}}

\EquationsNumberedThrough    

\TheoremsNumberedThrough     
\ECRepeatTheorems  %

\def\TITLEfont{\fs.18.22.\rm}
\makeatletter
\def\theARTICLETITLE{%
  \HOOKtop
  \begin{flushleft}
  \vspace*{-1cm}%
  \TITLEfont\HD{24}{0}\sffamily\bfseries\theTITLE\HD{0}{15}%
  \end{flushleft}}
\def\theARTICLEABSTRACT{%
  \begin{Center}
  \HOOKb
    \begin{minipage}[t]{\textwidth}\parindent1em
    \ABSfont
    \noindent\theABSTRACT\endgraf
    \vskip5pt
    \theFUNDING
    \theKEYWORDS
    \theSUBJECTCLASS
    \theAREAOFREVIEW
    \theMSCCLASS
    \theORMSCLASS
    \if@BLINDREV\else\theHISTORY\fi
  \noindent  \hrulefill
  \end{minipage}%
  \vspace*{0pt}
  \end{Center}}
\makeatother
\makeatletter
\RRHFirstLine{\bf\theRUNAUTHOR:\enskip {\it\theRUNTITLE}}%
\LRHFirstLine{\bf\theRUNAUTHOR:\enskip {\it\theRUNTITLE}}%
\RRHSecondLine{}%
\LRHSecondLine{}%
\def\setoddRH{\hbox to \textwidth{\fs.7.8.\tabcolsep0pt
  \begin{tabular*}{\textwidth}[b]{l@{\extracolsep\fill}r}
  {\theRRHFirstLine}& \raisebox{0pt}[0pt][0pt]{\fs.10.10.\thepage}\\[-4pt]
  \rlap{\VRHDW{0.5pt}{0pt}{\textwidth}}&\\
  \end{tabular*}}}
\def\setevenRH{\hbox to \textwidth{\fs.7.8.\tabcolsep0pt
  \begin{tabular*}{\textwidth}[b]{l@{\extracolsep\fill}r}
  \raisebox{0pt}[0pt][0pt]{\fs.10.10.\thepage}&{\theLRHFirstLine}\\[-4pt]
  \rlap{\VRHDW{0.5pt}{0pt}{\textwidth}}&\\
  \end{tabular*}}}
\makeatother

\begin{document}

\RUNAUTHOR{Allu and Kagan}

\RUNTITLE{Can AI Chat Replace Traditional Search?}
 
\TITLE{Can AI Chat Replace Traditional Search? Evidence from a RAG-based Live Events Experiment} 
\makeatletter
\if@BLINDREV
\long\def\ARTICLEAUTHORS#1{\gdef\theARTICLEAUTHORS{}}%
\LRHFirstLine{}\RRHFirstLine{}
\fi
\makeatother
\ARTICLEAUTHORS{%
\AUTHOR{Rakesh Allu}
\AFF{University of Illinois, Urbana Champaign, Gies College of Business}

\AUTHOR{Evgeny Kagan}
\AFF{Johns Hopkins University Carey Business School}
} 

\ABSTRACT{%
Conversational AI tools are being increasingly integrated into online search and browsing interfaces. To understand how such tools affect user behavior and search outcomes, we run a pre-registered, incentivized online experiment in which participants search a catalog of several thousand events scraped from a major live-events platform. We compare three interfaces: a classic, attribute-based filter interface (\textit{Filter}), a GenAI chat interface (\textit{Chat}), and a combined interface offering both (\textit{Hybrid}). In both AI conditions, AI output is restricted via retrieval-augmented generation (RAG). Our main result is that, relative to the other two conditions, the \textit{Chat} condition performs the worst. The effect is robust across AI models (\texttt{gpt-4.1-mini}, \texttt{gpt-5.2} and \texttt{gpt-5.4}) and  is driven mainly by users who perform open-ended exploration (as opposed to users searching for a specific artist, venue or sports team). Furthermore, there are no significant differences on any metrics between \textit{Filter} and \textit{Hybrid}. We also document two  mechanisms. First, \textit{Chat} creates an \textit{echo chamber}: without a visible menu, users construct queries based on what they think is contained in the catalog. As a result, they frequently search for event types that are not available in the catalog, and they search too narrowly. Second, \textit{Chat} leads to  \textit{prompting fatigue}: it shifts effort towards prompting, and away from inspecting and evaluating the output. As a result, users often pick one of the first displayed suggestions rather than examining the list more carefully. Taken together, our results suggest that AI is best used as a complementary tool to existing filter-based search rather than as a replacement.
}%


\maketitle
\vspace{0.2cm}

\section{Introduction}\label{sec:Intro}
\begin{figure}[tb]
\caption{Search interface of Ticketmaster.}
\label{fig:ticketmaster}
\centering
\includegraphics[width=\textwidth]{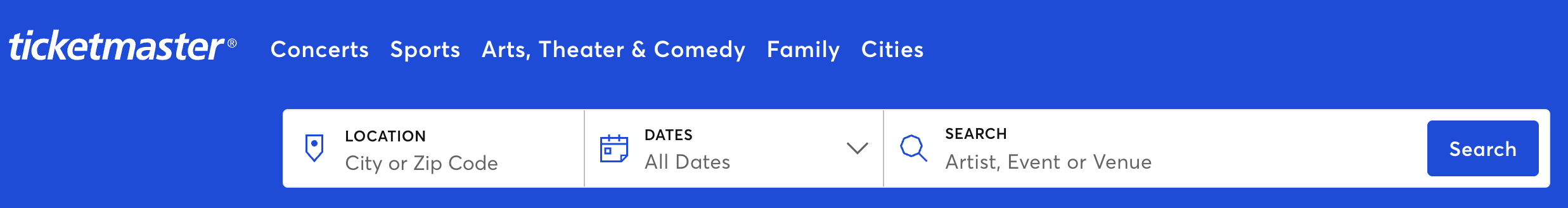}
\end{figure}

Digital platforms have traditionally helped consumers navigate large inventories of products and services through attribute-based filters and text-based keyword search. A typical example is the search page of Ticketmaster, a leading live-events platform (Figure~\ref{fig:ticketmaster}). The page is organized around category tabs, a location filter, a date filter, and a keyword box. To find an event users must translate their preferences into some combination of these predefined fields. More recently, platforms have started  experimenting with a different design, in which users type their queries in natural language. These conversational (AI-based) interfaces are built on generative language models that have become sufficiently capable for routine commercial use. The promise of conversational (AI-based) search is that it can accommodate open-ended queries and preferences that are difficult to express through pre-defined filters. As KAYAK's CEO Steve Hafner puts it, \textit{``Travelers aren't limited to preset entry fields anymore. Just tell us what you're looking for in plain language, and we'll answer the same way.''}\footnote{KAYAK AI search: \url{https://www.kayak.com/news/ai-mode/}} Indeed, KAYAK, Expedia,\footnote{Expedia AI search: \url{https://techcrunch.com/2024/05/14/expedia-starts-testing-ai-powered-features-for-search-and-travel-planning/}} and Booking.com\footnote{Booking.com AI features: \url{https://news.booking.com/bookingcom-enhances-travel-planning-with-new-ai-powered-features--for-easier-smarter-decisions/}} have all deployed or are actively piloting AI tools on their platforms.

Despite widespread enthusiasm among industry leaders, it is an open question whether AI adds value to consumer search and exploration, and how this value differs for specific use cases and implementations. On the one hand, the classic argument from the search literature \mbox{\citep{Stigler1961, Bakos1997, BrynjolfssonHuSmith2003}} is that a less-constrained search should produce greater value for the consumer. This suggests that AI should improve search outcomes as it relaxes the hard constraints of attribute-based filters and keyword-based search and lets users express moods and occasions that cannot be easily codified. For example, on an events platform, a chat user can simply ask for \emph{``a fun night out with friends''} or \emph{``something romantic for date night''}, which cannot be expressed through any combination of filter categories. On the other hand, AI may increase search frictions: forming and typing natural-language queries and iterating with the AI may demand more time and effort than using filters to select a subset of the available options. Indeed, a 2026 survey found that 58\% of US consumers are open to AI-assisted shopping but only 6\% have tried it, with conversational tools gaining traction only for simpler, repeat purchases \citep{Choudhary2026RetailBrew}. Thus, in this paper, we ask: 

\vspace{0.2cm}
\noindent \emph{(i) How does conversational AI affect search outcomes (value of the chosen alternative, search costs) relative to conventional search?} 
\vspace{0.2cm}

While it is easy to think of AI adoption as a binary decision, platforms must also decide \textit{how} AI should be integrated with existing search tools. Should AI be treated as a substitute for or a complement to conventional search? Having access to both modalities may be especially valuable when consumers can specify some preferences precisely but can express others only in broad or subjective terms. For example, filters may be well-suited to handle the codifiable part of search, such as a date or a genre, while AI can handle more open-ended preferences such as a mood or an occasion. This leads to our second, more practical question:
\vspace{0.2cm}

\noindent \emph{(ii) Should AI tools replace conventional search or be shown alongside it?}

\vspace{0.2cm}

The answer to (i) and (ii) may further depend on two factors. First, the value of conversational search may depend on the capability of the underlying AI model. A state-of-the-art model may be better able to synthesize the information in the retrieved listings and communicate this to the user than an older, less capable model. Second, search itself can range from targeted retrieval where the user knows exactly what they are looking for to broader, more open-ended browsing and exploration \citep{Marchionini2006, Moe2003}. For example, a user on a live-events platform may search for \textit{``NY Knicks''} or \textit{``Hamilton''}, whereas a more exploratory search might be \textit{``something fun to do this Friday''} or \textit{``a romantic date-night idea''}. We therefore ask: 
\vspace{0.2cm}

\noindent \emph{(iii) Does the value of conversational AI in search depend on the capabilities of the model? Does it depend on whether the user is performing targeted retrieval or broad exploration?}

\subsection{Research Design}
We address questions (i)-(iii) in the context of experiential goods such as live events, video streaming, travel and lodging, and dining. A defining feature of this setting is that it often includes an exploration stage where users need to browse the available options in order to understand the search space and identify their preferred alternatives \citep{Slovic1995,BettmanLucePayne1998, Moe2003}. Within the experiential goods category, we use a live-event ticketing platform (\url{ticketmaster.com}) as our empirical setting. 

We begin by compiling a catalog of several thousand events scheduled in New York City over a two-month window.\footnote{In particular, we scrape all Ticketmaster events with available tickets that occur in the two month period that starts approximately two weeks after the end of our experimental data collection. We choose NYC because it is the market with the largest inventory of events on Ticketmaster.} We then recruit New York and New Jersey residents to participate in the experiment. Participants' task is to search the inventory of events with the goal of selecting the event that they find most appealing. To make the choice consequential, incentives are tied to the search outcome: at the end of the study, a randomly selected subset of participants has a chance to receive real tickets to the event they chose, so every participant is incentivized to find an event they would actually attend. We measure how much each participant values their selected event using the Becker-DeGroot-Marschak (BDM) mechanism \citep{BeckerDeGrootMarschak1964}, which has the advantage of being incentive-compatible.\footnote{Under the BDM mechanism the dominant strategy is to report one's true valuation \citep[see also][for similar elicitation methods in consumer-choice experiments]{DingGrewalLiechty2005, LeeDonohue2026}.} Additional outcome measures capture the search process (time spent searching, number of iterations, interaction patterns) and self-reported satisfaction with both the search process and the selected event.

We conduct three between-subjects treatments. In the first treatment, which we will refer to as \textit{Filter}, participants use a classic, attribute-based interface with drop-downs for event date and type (Sports, Arts \& Theater, Music) and a search box. The search box performs simple text matching: it returns events whose text contains the exact string of characters entered by the user. This reproduces the dominant design used on most experiential good search platforms such as Ticketmaster and Eventbrite.\footnote{For representative examples of this interface design, see \url{https://www.ticketmaster.com}, \url{https://www.stubhub.com}, \url{https://www.seatgeek.com}, and \url{https://www.vividseats.com}.} In the second treatment, which we will refer to as \textit{Chat}, participants use a chat interface backed by a large language model (LLM) that has been given access to the same inventory of events as the filter-based search through Retrieval Augmented Generation (RAG), and that returns events with the closest match (``closest match'' will be defined later). Finally, in the third treatment, which we will refer to as \textit{Hybrid}, participants see the same interface as the \textit{Filter} treatment with the search box now powered by an LLM. That is, \textit{Hybrid} includes design elements of both \textit{Filter} and \textit{Chat} treatments. 

A key feature of our design is the RAG architecture used in the \textit{Chat} and \textit{Hybrid} conditions. RAG restricts the AI to respond only based on the event listings that exist in our database. This has two advantages. First, it ensures that all three experimental conditions draw from the same set of events. Prior to conducting the experiment, we verify that, for the same query, all our treatments produce the same list of results. This allows us to trace any potential differences in participants' behavior and outcomes to the interface and the retrieval process, rather than to differences in the event lists being displayed. Second, RAG reflects state-of-the-art deployment of AI-based conversational search in practice \citep{GaoEtAl2023, NVIDIA2023RAG, AWS2024RAG, SongGaurZhu2025}. Our experiment therefore reproduces the realistic deployment conditions of an AI-assisted search platform while maintaining experimental control. Furthermore, to test whether the results are caused by any particular model behavior, language or personality, we also vary the model used to translate RAG output into conversational output in the \textit{Chat} treatment. In particular, we test both an older, legacy model (\texttt{gpt-4.1-mini}), as well as two state-of-the-art models available during the data collection (\texttt{gpt-5.2} and \texttt{gpt-5.4}). 

\subsection{Preview of Results} Our main result is that the value of conversational AI search depends on its implementation. Regarding Research Question \textit{(i)}, we find that the treatments are ranked as follows in terms of search performance: $\textit{Hybrid} \approx \textit{Filter} > \textit{Chat}$. In particular, \textit{Chat} is significantly outperformed by both \textit{Hybrid} and \textit{Filter} on three of the four main outcome measures. Participants value their chosen event about 22\% more under \textit{Hybrid} than under \textit{Chat} and 12\% more under \textit{Filter} and than under \textit{Hybrid}. By contrast, \textit{Filter} and \textit{Hybrid} are statistically indistinguishable. Furthermore, the costs of search (time spent interacting with the system) are ranked as follows: $\textit{Chat} > \textit{Hybrid} \approx \textit{Filter}$. That is, \textit{Chat} is not only the lowest performing, but also the costliest (from the consumer's perspective) interface. On average, \textit{Chat} participants spend 26\% more time in a session than those in \textit{Filter} or \textit{Hybrid}. Their ratings of both the search process and of the search outcome are also significantly lower than those of the other two treatments. 

The treatment comparisons together answer our Research Question \textit{(ii):} using conversational tools in tandem with conventional tools (our \textit{Hybrid} arm) is close to costless, whereas replacing the filter with chat lengthens search and lowers search performance. Furthermore, with respect to Research Question \textit{(iii)}, we find that search intent is a moderator of the treatment effects. In particular, users performing open-ended exploration value their choices significantly less in the \textit{Chat} arm relative to the other two arms, whereas users performing retrieval (looking for a specific artist, venue, or team) do just as well under \textit{Chat} as they do in the other conditions. Finally, our results do not depend on \emph{which} model powers the chat: whether the responses are generated by our legacy model (\texttt{gpt-4.1-mini}), or by newer models  (\texttt{gpt-5.2} and \texttt{gpt-5.4}) does not significantly predict any of the outcome variables. 

Why does the \textit{Chat} condition perform poorly, while the \textit{Hybrid} condition performs on par with the traditional filter-based search? We trace this back to two mechanisms: the \textit{echo chamber} effect and the \textit{prompting fatigue} effect. The \textit{echo chamber} effect occurs because \textit{Chat} users, unlike those in \textit{Filter} and \textit{Hybrid}, do not see a menu of available event types. They must instead compose queries from what they believe is available in the catalog, and the model reflects those beliefs back in the form of search results. Search results are therefore limited to the part of the catalog the user already had in mind, or the user finds that the searched events are not in the catalog at all. Searching too narrowly and searching out-of-catalog are both associated with lower valuations. The \textit{prompting fatigue} effect occurs because the \textit{Chat} interface redirects the user's effort and attention away from carefully examining the retrieved events and towards constructing effective queries and waiting for the system to return a list of events. This causes users to pick results near the top of the list, which in turn is associated with lower valuations. Together, the mechanisms suggest that structured interfaces provide consumers with a map of the feasible choice space, whereas a blank conversational interface makes users construct that choice space themselves, which can lead to poor search outcomes.

\subsection{Contributions}\label{subsec:contributions}

We make three contributions: 

\begin{itemize}
    \item [(i)] \emph{Empirically}, we provide evidence that adding a conversational LLM to a filter-based interface does not affect search costs or search performance, while removing structured filters can leave consumers worse off, particularly when they explore broadly rather than search for a specific target. This adds to the behavioral literature on human-AI interactions, which largely studies AI agents that provide advice \citep[e.g.,][]{DietvorstSimmonsMassey2015, LoggMinsonMoore2019}, perform customer service \citep[e.g.,][]{LuoTongFangQu2019, AdamWesselBenlian2021, SongGaurZhu2025, KaganHathawayDada2026} or generate product or price comparisons \citep{SpathariotiEtAl2025, LinShiSun2025}. In contrast, we study AI as the interface through which consumers search. 
    \item [(ii)] \emph{Practically}, our results offer guidance for deploying AI as a search tool. Because filters and chat have complementary strengths, and because even the lightweight, legacy \texttt{gpt-4.1-mini} performs on par with frontier models, platforms are better advised to invest in the design of the search interface rather than in more capable, and more costly, AI models. Furthermore, we show that BDM valuations and survey-based satisfaction scores agree on most measures. Platforms (for whom satisfaction surveys are standard practice while BDM-style elicitation is not) can thus evaluate the costs and benefits of deploying AI tools from survey data alone. 
    \item [(iii)] \emph{Methodologically}, we develop the first (to our knowledge) behavioral experiment that leverages retrieval-augmented generation (RAG). The advantage of RAG is that it reconciles realism with experimental control in studying human-AI interactions. A RAG-based system ensures that output is verifiably consistent across model runs and across users given the same inputs (as opposed to using an LLM system prompt, which may produce stochastic output even for identical inputs). Our work thus provides a template for future experimental studies, particularly ones that involve information retrieval.  
\end{itemize}

\section{Hypothesis Development and Literature}\label{sec:Literature}
We begin by reviewing the literature that is most closely related to our research questions and use it to motivate our hypotheses (\textsection\ref{subsec:hypotheses}). Additionally, we survey different LLM customization approaches because a key component of our experimental approach is a customized LLM (RAG) (\textsection\ref{subsec:llm-customization}), and discuss broader behavioral work related to human-AI interactions (\textsection\ref{subsec:human-ai}).

\subsection{Hypothesis Development}\label{subsec:hypotheses}
We formulate hypotheses related to two outcomes: the value that conversational AI adds to consumer search and the cost at which that value is obtained. 

\subsubsection{Value of Search}\label{subsec:search-rec-systems}
To measure the value of search, we use the participant's valuation (also sometimes called ``willingness to pay'' in the literature) of their chosen event, elicited via the Becker-DeGroot-Marschak (BDM) mechanism \citep{BeckerDeGrootMarschak1964}. See \cite{WertenbrochSkiera2002} and \cite{DingGrewalLiechty2005} for the validation of this method in consumer choice experiments. In addition, we elicit users' satisfaction with the chosen event on a 7-point Likert scale. Prior work suggests that monetary valuation and subjective evaluations capture related but distinct aspects of user preferences \citep{homburg2005satisfied}. Collecting both measures thus gives us a fuller picture of the value of search in each interface. 

Studies in economics and marketing have examined consumer choice in the presence of decision aids. Classic economic reasoning suggests that AI should be helpful. This is because search is costly \citep{Stigler1961, Weitzman1979} and consumers typically evaluate only a limited consideration set \citep{HauserWernerfelt1990}. Tools that lower search cost and widen the set of options should thus increase consumer welfare \citep{Bakos1997, HaublTrifts2000}. Since any query that can be posed through a filter can also be expressed in natural language (but not vice versa), the \textit{Chat} interface should be more helpful at finding relevant events. The gains should be largest in the long tail of niche options, which may be difficult to reach with filters. Indeed, earlier technologies that lowered search costs, from online catalogs to search and recommendation tools, shifted consumption toward niche products and created substantial consumer surplus simply by making more of the catalog reachable \citep{BrynjolfssonHuSmith2003, BrynjolfssonHuSimester2011}. Furthermore, our \textit{Hybrid} treatment, which offers both filter and chat modalities, should further increase the set of reachable events by allowing users to choose the modality (filters, chat or both) that they deem more useful or easy to use. These arguments motivate our first pre-registered hypothesis:\\
\vspace{-0.5cm}

\noindent\textbf{H1:} Valuations and satisfaction with the chosen event are ranked as $\textit{Hybrid} > \textit{Chat} > \textit{Filter}$.
\vspace{0.2cm}

However, some prior work also speaks against H1. \citet{CasteloBosLehmann2019} show that consumers trust algorithms for objective tasks with a verifiable answer but resist them for subjective, taste-based tasks. \citet{LongoniCian2022} formalize this as the \emph{word-of-machine} effect: consumers prefer AI for utilitarian attributes and humans for hedonic, experiential ones \citep[also see][]{LongoniBonezziMorewedge2019, PuntoniEtAl2021}. Live events are hedonic and experiential \citep{BatraAhtola1990, DharWertenbroch2000}, so this literature would predict that AI may perform poorly in our setting. 

\subsubsection{Cost of Search}\label{subsec:interfaces}
We measure search costs through the time the user spends interacting with the system. Search theory treats time as the main resource that people give up when searching \citep{Stigler1961, Becker1965}, and empirical work shows that time and effort costs differ across channels \citep[e.g.,][]{BrynjolfssonSmith2000, HannTerwiesch2003, ChintaguntaChuCebollada2012}. In addition to measuring time spent, we capture users' subjective perception of search costs by asking them to evaluate their satisfaction with the search process on a 7-point Likert scale. We add this measure both because perceived duration may differ from actual duration \citep{AntonidesVerhoefVanAalst2002} and because the evaluation of the process may depend on the content of the wait rather than its duration alone \citep[][]{maister1984,BuellNorton2011}.

A sizable human-computer interaction (HCI) and information retrieval literature studies how the design of a search interface affects search effort. Conversational search starts with an empty box, so the user must recall what they want and put it into words \citep{RadlinskiCraswell2017}. The process of composing a natural-language query has been shown to be effortful relative to menu choices \citep{ZamfirescuPereiraEtAl2023, SubramonyamEtAl2024}. By contrast, filter-based search requires a single selection from a drop-down menu which requires minimal effort. Furthermore, the \textit{Hybrid} treatment falls in between because composing text is optional rather than being required. Users can perform most of their search through low-effort menu selections and incur the cost of formulating a query only when they expect it to be worth the effort. We therefore hypothesize:

\smallskip

\noindent\textbf{H2:} Search time is ranked $\textit{Filter} < \textit{Hybrid} < \textit{Chat}$. Satisfaction with the search process is ranked $\textit{Filter} > \textit{Hybrid} > \textit{Chat}$.
\smallskip

\noindent A competing view is that conversational AI \emph{reduces} search costs rather than increasing them since the model requires fewer iterations to get to a good set of options \citep{Bakos1997, HaublTrifts2000}. Whether the added cost of composing a query is offset by a better match (produced by a given query), which may lead to fewer queries per user, is ultimately an empirical question and is one of the questions that we address in this paper. 


\subsection{Customized LLM Models}\label{subsec:llm-customization}
There are three main approaches that researchers have used to build custom LLMs: system prompting, model fine-tuning and RAG models. A set of representative studies using each approach and the corresponding domains in which they are used are summarized in Table~\ref{tab:approaches-by-task}. Most studies that examine how humans interact with LLMs rely on system prompting to augment the LLM. This approach has been used in experimental settings such as judgment and decision-making \citep[e.g.,][]{ChenKirshnerEtAl2024, ChenKirshnerEtAl2025, BinzSchulz2023, Horton2023, HagendorffEtAl2023, SuriEtAl2024}, market research \citep{BrandIsraeliNgwe2023, WangZhangZhang2025}, content and idea generation \citep{ChenChan2024, GirotraEtAl2023} and workplace productivity \citep{NoyZhang2023, DellAcquaEtAl2023, PengEtAl2023}. Fewer studies use model fine-tuning: \citet{BrynjolfssonLiRaymond2025} and \citet{ReisenbichlerEtAl2022} examine the effects of fine-tuned models in customer-service conversations and marketing content, respectively. The only RAG-based study that we are aware of is \cite{SongGaurZhu2025}, a field experiment in the customer service domain. 

In consumer search, the setting closest to ours, \cite{SpathariotiEtAl2025} study how LLM mistakes affect consumers' product purchase decisions. Their LLM generates responses from parametric memory, so it can hallucinate on factual queries; the authors build their task around an item the model reliably gets wrong to show that users accept its answer even when it is incorrect. Relatedly, \cite{LinShiSun2025} study how an AI chatbot affects online shopping on an Amazon-style platform. In their setting, the AI is a summarization and comparison technology (rather than a search tool) which helps summarize product descriptions, so the relevant context is small and can be passed directly in the prompt. In contrast to these studies on consumer search, our study \textit{requires} a RAG approach because our chatbot needs to draw on a large catalog of heterogeneous events and return a set strictly within the catalog that best matches the query. In particular, our study is the first behavioral experiment that we are aware of to use the RAG approach. In \textsection\ref{app:rag} we provide a more detailed comparison of the three approaches to building custom LLMs.

\begin{table}[tb]
\TABLE
{LLM Customization in the Experimental Literature.\label{tab:approaches-by-task}}
{\footnotesize
\begin{tabular}{l L{9.5cm} l}
\toprule
\textbf{Task/Domain} & \textbf{Experimental Studies} & \textbf{Approach} \\
\midrule
Judgment \& decision-making & \citet{ChenKirshnerEtAl2024, ChenKirshnerEtAl2025}; \citet{BinzSchulz2023}; \citet{Horton2023}; \citet{HagendorffEtAl2023}; \citet{SuriEtAl2024} & Prompting \\ [1ex]
Market research             & \citet{BrandIsraeliNgwe2023}                       & Prompting, fine-tuning \\
                            & \citet{WangZhangZhang2025}                         & Prompting              \\ [1ex]
Content and idea generation & \citet{ChenChan2024}; \citet{GirotraEtAl2023}      & Prompting              \\
                            & \citet{ReisenbichlerEtAl2022}                      & Fine-tuning            \\ [1ex]
Workplace productivity      & \citet{NoyZhang2023}; \citet{DellAcquaEtAl2023}; \citet{PengEtAl2023} & Prompting \\ [1ex]
Customer service            & \citet{BrynjolfssonLiRaymond2025}                  & Fine-tuning            \\
                            & \citet{SongGaurZhu2025}                            & RAG                    \\ [1ex]
Product search \& selection & \citet{SpathariotiEtAl2025}; \citet{LinShiSun2025} & Prompting              \\
                            & \textbf{This paper}                                         & RAG                    \\
\bottomrule
\end{tabular}}
{}
\end{table}

\subsection{Human-AI Interactions Beyond Consumer Search}\label{subsec:human-ai}
An extensive behavioral literature studies how people respond to algorithmic advice. Typically, in this literature algorithmic behavior is set by the experimenter and delivered in a fixed, structured form, e.g., a numeric estimate or a short recommendation. Findings range from \emph{algorithm aversion}, in which people discount algorithmic advice after seeing the algorithm err even when it outperforms humans \citep{DietvorstSimmonsMassey2015}, to \emph{algorithm appreciation} in other settings \citep{LoggMinsonMoore2019}. This literature also examines factors that determine users' uptake of AI-advice. For example, \citet{BalakrishnanFerreiraTong2026} find that decision-makers with private information weigh AI advice naively;   \citet{DiSorboFerreiraBalakrishnanTong2025} find that warning and endorsement cues help decision makers adjust AI forecasts for outliers, and \citet{LongoniCian2022} find that the uptake of advice is domain-dependent.

A separate stream of literature studies the deployment of conversational AI in customer support settings. In settings where AI acts as an aid to the customer service agent, \citet{BrynjolfssonLiRaymond2025,
ZhangNarayandas2025, NiEtAl2026} find that GenAI suggestions improve agent performance and customer sentiment in both B2B and B2C settings. In settings where AI is customer facing, \citet{LuoTongFangQu2019} find that a chatbot sells as well as a professional human agent but purchases fall once it discloses being a
machine, \citet{AdamWesselBenlian2021} find that anthropomorphic design cues raise task compliance among consumers to chatbot requests, and \citet{SongGaurZhu2025} find that LLM chatbots outperform rule-based chatbots. \citet{KaganHathawayDada2026} document the sources of aversion to customer service chatbots. Different from this work, where the AI acts as an advisor or a service agent whose output (a forecast, a recommendation, or a resolution) is presented to a human decision-maker, we study AI as a tool that helps the user perform search and exploration.

\section{Experiment Design}\label{sec:Methodology}
To study the effects of conversational search tools on search behavior and search performance (value obtained by the user), we conduct an online experiment in which participants are asked to search an inventory of live events and select the one they find most appealing. This section introduces our experimental methodology and treatments. 

\subsection{Sample and Exclusions}\label{sec:meth-overview}
A total of $389$ participants were recruited through CloudResearch Connect, with the recruitment pool being restricted to adults residing in New York or New Jersey.\footnote{Our pre-registration, which describes the subject pool, recruitment criteria, sample sizes and hypotheses, is available at \url{https://aspredicted.org/zm499g.pdf}. We used CloudResearch Connect, because this platform allows nonmonetary incentives as well as allows communication with participants outside of the experiment, which was necessary to deliver event tickets to selected participants. Many other platforms (e.g., Prolific) do not allow participants to receive non-monetary gifts outside of the platform's payment system. We had originally planned to recruit New York residents only, but the CloudResearch pool of NY-based subjects was too small to reach our pre-registered sample sizes. We therefore extended recruitment to New Jersey residents (Residence is controlled for in the analysis).} Participants were required to use a desktop device to complete the study. We applied three pre-registered exclusion criteria: (i) participants who did not return a valid CloudResearch completion code (5 participants), (ii) participants who answered four or more comprehension-quiz questions incorrectly (12 participants), and (iii) participants whose total time spent on the task was less than a third of the median (2.7 minutes, 12 participants), or more than three times the median (24.1 minutes, 12 participants). The median session length in the analysis sample was 8 minutes (average: 9.3 minutes). Because the criteria are not mutually exclusive, 27 unique participants were excluded. This yields a final sample of $362$ participants ($n = 104$ in \textit{Filter}, $n = 156$ in \textit{Chat}, and $n = 102$ in \textit{Hybrid}), which is approximately equal to the pre-registered sample sizes of 100, 150 and 100, respectively. The mean age in the analysis sample is 37.8 years (median 35), 53\% of participants identify as female, 31\% hold a four-year college degree or higher, and  79\% (resp.: 21\%) reside in New York (resp.: New Jersey). Balance tests are reported in \textsection\ref{app:balance}. Participants receive a fixed payment of \$4.00, plus a potential payment/prize based on the BDM mechanism (as will be explained in \textsection \ref{sec:meth-procedure}). 

\subsection{Event Catalog and AI Models}\label{sec:meth-catalog}
Our search inventory includes all live events scheduled in NYC over the course of two months. All events were scraped from Ticketmaster. Each event is characterized by a set of attributes provided by Ticketmaster: name, date, start time, venue, NYC borough, a segment which takes three values (\emph{Music}, \emph{Sports}, or \emph{Arts \& Theater}) and event description (genre and subgenre). We drop all free events and keep only events that have available tickets priced below \$200.\footnote{The cap limits our cost of fulfilling tickets to BDM lottery winners.}

Our choice of the empirical setting has three advantages. First, Ticketmaster is the largest live-events platform in North America, and New York City has the largest event inventory on the platform. Therefore, the set of events in the catalog is large enough to allow extensive browsing and exploration, which is the behavior that we want to study. Second, the set of events is heterogeneous enough that learning its contents requires effort, making search aids such as \textit{Filter} or \textit{Chat} consequential (the catalog includes 725 distinct event names and 136 venues, and spans 42 genres and 107 subgenres). Third, our pilots (\textsection\ref{app:pilots}) confirmed that the NY/NJ region provides a sufficiently large number of eligible participants on CloudResearch Connect and that the majority of the subjects live within reasonable travel distance to most venues.

The experiment was conducted in two waves. Wave 1 was run in March 2026 and was based on a catalog of NYC events scheduled over April and May 2026 (4,066 events total). Wave 2 was run in April 2026 with the catalog refreshed to cover May and June 2026 (3,836 events total). In both waves, participants in the \textit{Chat} condition were randomly assigned to one of three AI models: \texttt{gpt-4.1-mini}, \texttt{gpt-5.2}, or \texttt{gpt-5.4}. The retrieval pipeline (embedding model, FAISS index, similarity cutoff, temporal pre-filter, retrieval cap, router LLM, and system prompts) was held fixed across both waves and across all three model conditions; only the AI model varied. The two waves of data collection and the three AI models allow us to test whether search outcomes depend on the performance of the AI model powering the chat, or on the interaction between a given model and the particular set of event listings being searched.

\begin{figure}[p]
\caption{Search Interface in Each Treatment}
\label{fig:screens-initial}
\centering
\vspace{-0.6cm}
\begin{subfigure}{0.65\linewidth}
\caption{\textit{Filter} Treatment}
\centering
\vspace{-0.2cm}
\includegraphics[width=\linewidth]{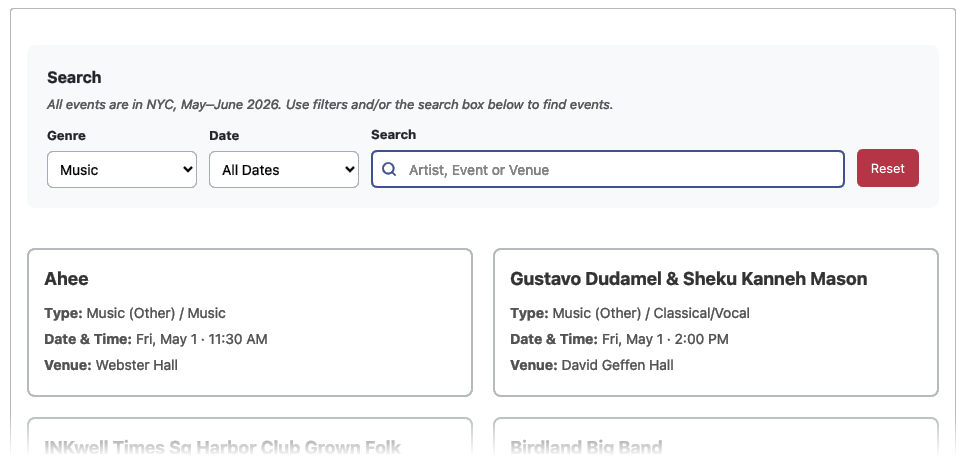}
\label{fig:screen-trad}
\end{subfigure}\\[12pt]
\begin{subfigure}{0.65\linewidth}
\centering
\vspace{-0.6cm}
\caption{\textit{Chat} Treatment}
\vspace{-0.2cm}
\includegraphics[width=\linewidth]{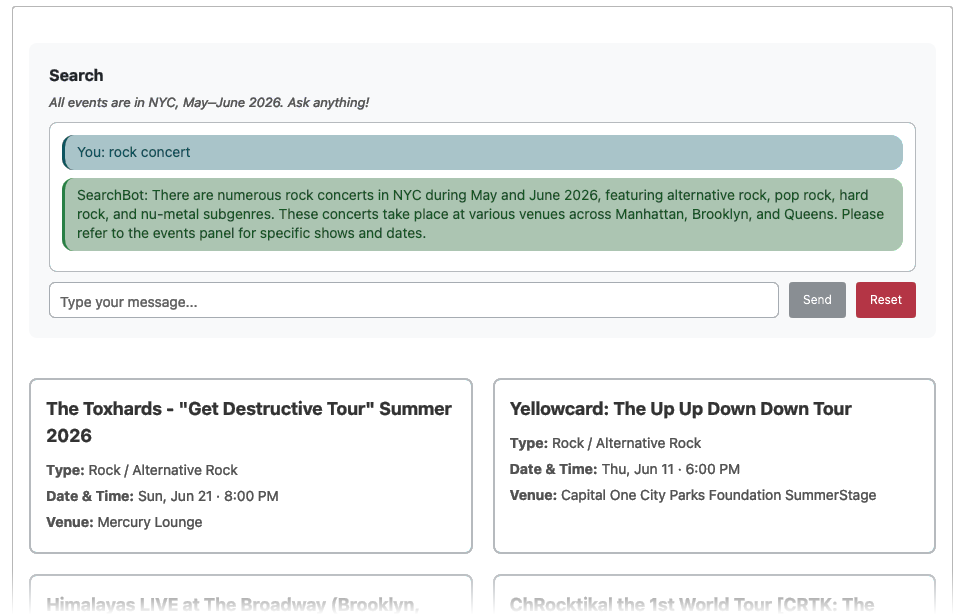}
\label{fig:screen-genai}
\end{subfigure}\\[12pt]
\begin{subfigure}{0.65\linewidth}
\centering
\vspace{-0.6cm}
\caption{\textit{Hybrid} Treatment}
\vspace{-0.2cm}
\includegraphics[width=\linewidth]{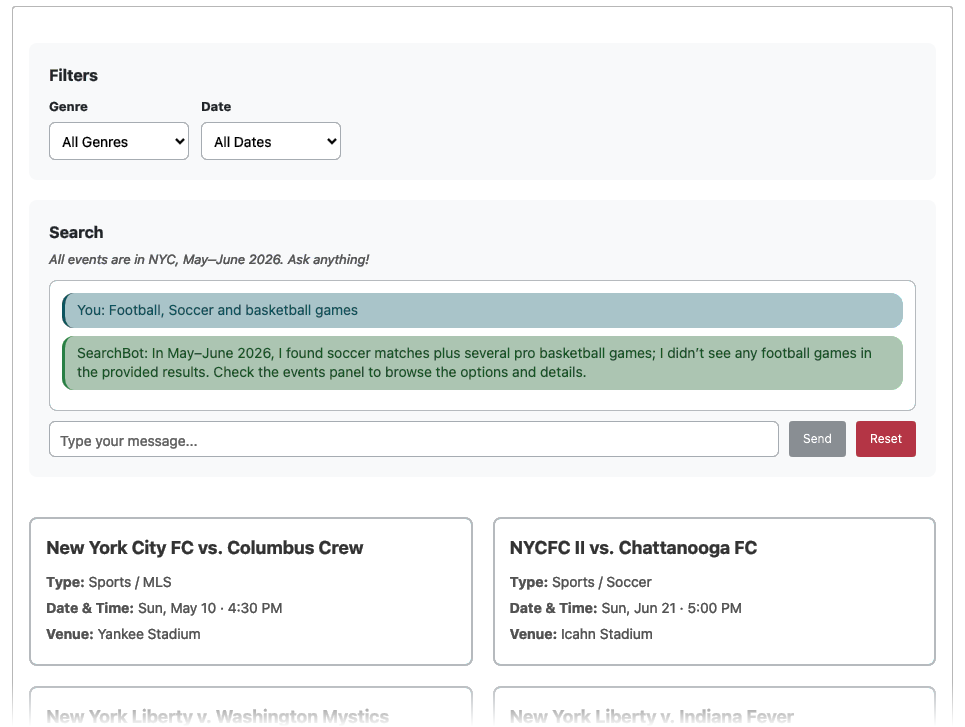}
\label{fig:screen-hybrid}
\end{subfigure}
\end{figure}

We administered three between-subjects treatments, implemented as an oTree-hosted web application \citep{chen2016otree}.\footnote{The experiment design was first tested in two pilots ($N = 72$ in total), both run on CloudResearch with NY/NJ residents and based on the same Ticketmaster catalog as the main experiment. \textsection\ref{app:pilots} provides an overview of the pilots and summarizes the main takeaways. The pilots were used to validate the BDM and satisfaction scales and to examine potential issues with the conversational interface. Neither pilot is included in the final analysis sample.} Figure~\ref{fig:screens-initial} shows the search page in each treatment.

\begin{itemize}
\item{\textit{Filter} Treatment. In this treatment participants are presented with a structured filter panel and a paginated results list. Filters include drop-down menus for date, segment, as well as a text input for free-text search across event names and venues. The search box performs simple text matching and returns events whose text contains the exact string of characters entered by the user. The results list shows ten events per page. This interface mirrors exactly the design currently used on real-world experiential-good platforms, such as \url{https://www.ticketmaster.com}, \url{https://www.stubhub.com}, \url{https://www.seatgeek.com}, and \url{https://www.vividseats.com}. Displayed results are ordered chronologically (by event date, then start time) and then alphabetically.} See Figure~\ref{fig:screens-initial}a for a screenshot.

\item{\textit{Chat} Treatment.} In this treatment participants are presented with a chat window. They can type queries in natural language and receive responses from a large language model that accesses the catalog through retrieval-augmented generation (\textsection\ref{sec:meth-rag}). Each response includes a short natural-language summary and a list of recommended events drawn from the catalog, paginated at ten events per page as in the \textit{Filter} treatment. The interface contains no structured filter panel: all interaction with the catalog occurs through the chat. Displayed results are ordered by $L_2$ distance and, for events with identical $L_2$ distances, chronologically (see \textsection\ref{sec:meth-rag} for details). See Figure~\ref{fig:screens-initial}b for a screenshot.

\item{\textit{Hybrid} Treatment.} In this treatment participants are presented with the structured filter panel from \textit{Filter} and the chat window from \textit{Chat} side by side. Participants are free to use either or both modalities at any point in the session. See Figure~\ref{fig:screens-initial}c for a screenshot.
\end{itemize}

\subsection{Retrieval-Augmented Generation (RAG)}\label{sec:meth-rag}
A key feature of our design is that the system in the \textit{Chat} and \textit{Hybrid} conditions can recommend only events from our catalog. To implement this restriction we use retrieval-augmented generation \citep{LewisEtAl2020RAG, GaoEtAl2023}, wherein the LLM is paired with a fixed \emph{knowledge base}. RAG is the standard way commercial platforms keep an AI assistant within the scope of the data \citep{SongGaurZhu2025}, and it has been shown to substantially reduce factual hallucination relative to general-purpose AI tools that rely on the model's training data alone \citep{ShusterEtAl2021}. 

\begin{figure}[h]
\caption{RAG Architecture in the \textit{Chat} and \textit{Hybrid} Treatments.}
\label{fig:rag-overview}
\centering
\smallskip
\begin{tikzpicture}[
  node distance=1.7cm,
  box/.style={draw, rectangle, rounded corners=2pt, minimum width=2.6cm, minimum height=1.0cm, align=center, font=\small\linespread{0.85}\selectfont},
  kb/.style={draw, rectangle, rounded corners=2pt, minimum width=2.8cm, minimum height=1.0cm, align=center, font=\small\linespread{0.85}\selectfont, fill=gray!12},
  arr/.style={-Latex, thick}
]
\node[box] (user) {User};
\node[box, right=of user, font=\scriptsize] (app)  {oTree\\application};
\node[kb, right=of app, font=\scriptsize]  (kb)   {Knowledge base\\(event catalog,\\vector store)};
\node[box, right=of kb, font=\scriptsize]   (llm)  {LLM};
\draw[arr] (user) -- node[font=\scriptsize, above] {message} (app);
\draw[arr] (app)  -- node[font=\scriptsize, above] {query} (kb);
\draw[arr] (kb)   -- node[font=\scriptsize, above, align=center] {events ranked\\by distance} (llm);
\draw[arr] (llm) to[bend right=32] node[font=\scriptsize, above] {LLM response + relevant event IDs} (app);
\draw[arr] (app) to[out=-130, in=-50, looseness=0.9] node[font=\scriptsize, below] {LLM response + event list} (user);
\end{tikzpicture}
\end{figure}
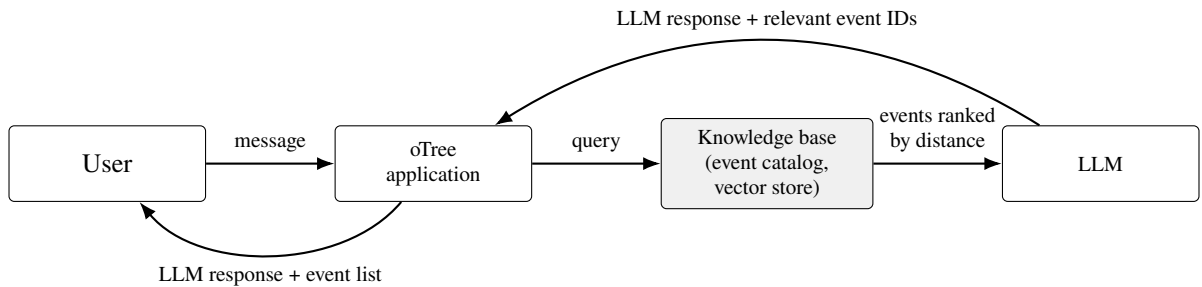

Figure~\ref{fig:rag-overview} shows how the RAG architecture is implemented in our experiment. When the user sends a message, a router LLM turns it, together with the recent conversation history, into a short search query. The backend retrieves the nearest neighbors to this query based on $L_2$ distance from the knowledge base (our scraped list of events with event details) and ranks them, again based on $L_2$ distance.\footnote{Nearest-neighbor retrieval by $L_2$ distance over embedding vectors is the standard approach in dense retrieval \citep{johnson2019billion}. Because the embedding vectors are unit-normalized (\textsection\ref{app:rag-pipeline}), ranking by $L_2$ distance is equivalent to ranking by cosine similarity.} The generation LLM sees these events and their relative ranking and writes a response based on that information; the oTree app then displays to the user the AI-generated response as well as the ranked event list. See Figure~\ref{fig:rag-overview} for the process flow and Figure~\ref{fig:embed-space} in \textsection\ref{app:rag-embedding} for an illustration of how retrieval works using two example queries. Within the \textit{Chat} treatment, routing and retrieval are identical across the three model conditions (\texttt{gpt-4.1-mini}, \texttt{gpt-5.2}, and \texttt{gpt-5.4}); only the generation step varies, so all three models see the same retrieved events and differ only in the wording of the reply.

The main advantage of our RAG-based design is that it combines realism with experimental control. In particular, it reproduces the open-ended, conversational search that consumers face when interacting with commercial GenAI tools while also holding the search results fixed and consistent across treatments and model runs. To further verify the latter claim, we performed a benchmark test in which we examined the responses to 1,000 queries per wave across all three treatments. In particular, we performed a Monte-Carlo simulation wherein we picked a date and an event name at random (from our database) and used that information to construct a query and test it in each of our treatments. The simulation confirmed that the returned listings are accurate and consistent across all treatments for over 95\% of the simulated queries (see \textsection\ref{app:rag-mc} for details). Thus, any potential differences in search performance between treatment arms can be attributed to how consumers respond to the interface, rather than to stochasticity or hallucination in the underlying model.

\subsection{Experimental Flow,  Measures \& Variables}
\label{sec:meth-procedure}
Figure~\ref{fig:design} summarizes the experimental protocol. After consenting, subjects were randomly assigned to one of three treatments. They then completed the main experimental task, i.e., finding the event that is most attractive to them, in their assigned treatment. There was no limit on the number of queries they could make or the time they could spend searching; the search ended when they selected an event. After that, we elicited their valuation of the selected event, which was followed by survey questions. \textsection\ref{app:materials} reproduces the instructions and comprehension questions.  

\begin{figure}[b]
\FIGURE
{\resizebox{\textwidth}{!}{%
\begin{tikzpicture}[
  >=Latex,
  box/.style={draw, rectangle, rounded corners=2pt, minimum width=1.9cm,
              minimum height=0.9cm, align=center, font=\small\linespread{0.9}\selectfont},
  rnd/.style={draw, diamond, aspect=1, align=center, inner sep=0.5pt,
              font=\footnotesize\linespread{0.9}\selectfont},
  arm/.style={draw, rectangle, rounded corners=2pt, minimum width=2.8cm,
              minimum height=0.9cm, align=center, font=\small\linespread{0.8}\selectfont},
  mbox/.style={draw, rectangle, rounded corners=2pt, minimum width=2.35cm,
               minimum height=0.7cm, align=center, font=\footnotesize\selectfont},
  arr/.style={-Latex, thick},
]
\node[box] (consent) {Consent};
\node[rnd, right=0.45cm of consent] (rand) {Random\\assignment};
\node[arm, right=0.6cm of rand] (gen)  {\textit{Chat} treatment};
\node[arm, above=1.5cm of gen]  (trad) {\textit{Filter} treatment};
\node[arm, below=1.5cm of gen]  (both) {\textit{Hybrid} treatment \\ (\texttt{gpt-5.4})};
\node[rnd, right=0.45cm of gen] (rand2) {Random\\assignment};
\node[mbox, right=0.45cm of rand2] (m2) {\texttt{gpt-5.2}};
\node[mbox, above=0.28cm of m2] (m1) {\texttt{gpt-4.1-mini}};
\node[mbox, below=0.28cm of m2] (m3) {\texttt{gpt-5.4}};
\node[box, right=0.7cm of m2] (bdm) {BDM\\valuation};
\node[box, right=0.5cm of bdm] (survey) {Demographics and\\satisfaction survey};
\draw[arr] (consent) -- (rand);
\draw[arr] (rand.east) -- (trad.west);
\draw[arr] (rand.east) -- (gen.west);
\draw[arr] (rand.east) -- (both.west);
\draw[arr] (gen.east) -- (rand2.west);
\draw[arr] (rand2.east) -- (m1.west);
\draw[arr] (rand2.east) -- (m2.west);
\draw[arr] (rand2.east) -- (m3.west);
\draw[arr] (m1.east) -- (bdm.west);
\draw[arr] (m2.east) -- (bdm.west);
\draw[arr] (m3.east) -- (bdm.west);
\draw[arr] (trad.east) -| (bdm.north);
\draw[arr] (both.east) -| (bdm.south);
\draw[arr] (bdm) -- (survey);
\path ([shift={(-0.5,-0.5)}]current bounding box.south west)
      ([shift={(0.5,0.5)}]current bounding box.north east);
\end{tikzpicture}%
}}
{Experimental Flow and Treatment Assignment\label{fig:design}}
{}
\end{figure}

Our primary outcome is the participant's \emph{valuation} of the chosen event, elicited via the BDM mechanism \citep{BeckerDeGrootMarschak1964}, implemented as a multiple price list (Figure~\ref{fig:mpl} shows the elicitation screen). We use a multiple-price-list implementation rather than asking participants to enter a single dollar amount because the format is known to be easier to understand and to produce more reliable estimates than direct elicitation \citep{HoltLaury2002, AndersenHarrisonLauRutstrom2006, CasonPlott2014}; see also \citet{DingGrewalLiechty2005, LeeDonohue2026} for related incentive-compatible elicitation in consumer-choice experiments. The price list ran from \$0 to \$50 in \$5 intervals. One in 20 participants was selected to receive the prize (money or a ticket to the event they selected, depending on their choice in the randomly selected row of the price list). See \textsection\ref{app:pilots} for the description of the pilots in which we validated the lower/upper bound used in the BDM elicitation.

We collect three additional outcome measures: the time spent searching (as a measure of search costs), as well as two satisfaction measures collected via post-task survey, both on a 1--7 Likert scale: \emph{process satisfaction} (with the search experience) and \emph{outcome satisfaction} (with the chosen event). Additionally, participants are asked about their search intent (whether they had a specific event in mind, or not), which will give us more insight into the usefulness of conversational search for different types of search processes. We also log search process data, including the content and timestamp of every filter change, text search, chat messages, and the events displayed in each result set. We use these logs to perform mechanism analyses in \textsection\ref{sec:Mechanisms}. 


\section{Results}\label{sec:results}
We begin by presenting summary statistics for the four main outcome measures introduced in \textsection\ref{sec:meth-procedure}. After that we test H1 and H2. For H1, we focus on examining treatment effects on the BDM valuation of the chosen event and on self-reported satisfaction with the chosen event. For H2, we focus on examining treatment effects on the total time spent searching and on the self-reported satisfaction with the search process.\footnote{Our pre-registration also lists other process measures: the number of distinct queries (filter changes or chat messages), number of words typed, number of resets, and the number of events viewed. We present all variables in Table~\ref{tab:process} in \textsection\ref{app:process-full}.}

\subsection{Summary Statistics}\label{sec:res-overview}
\begin{figure}[tb!]
\FIGURE
{\includegraphics[width=\linewidth]{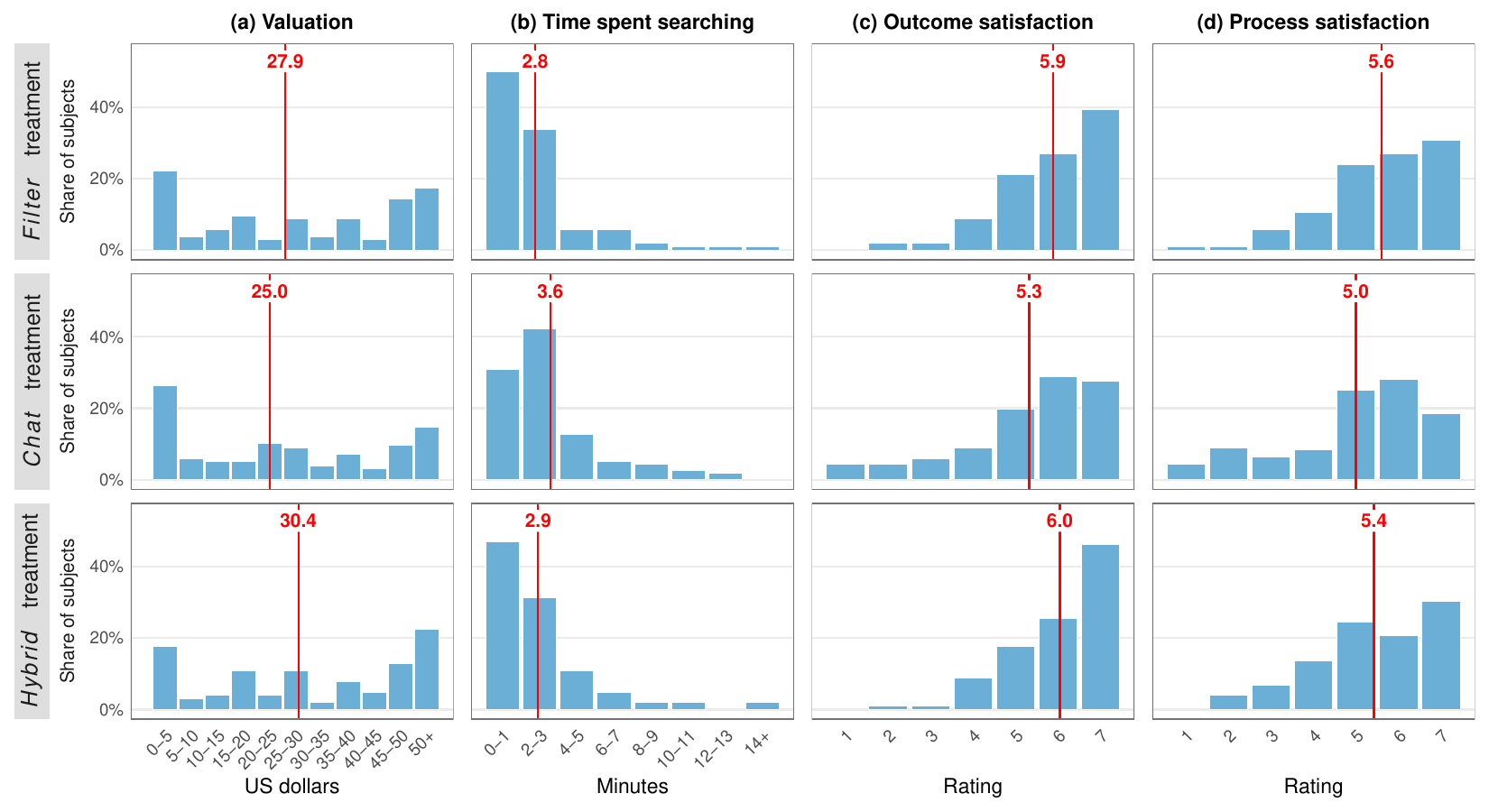}}
{Distribution of each primary outcome by treatment arm\label{fig:overview-dist}}
{\begin{minipage}[t]{\textwidth}\footnotesize\linespread{0.9}\selectfont The red vertical line in each panel indicates the treatment mean. In panel (a), to compute the mean, we use the midpoint of each \$5 bucket, with participants who never accepted cash (always preferred the event over cash) entering as \$52.50.\end{minipage}}
\end{figure}

Figure~\ref{fig:overview-dist} shows the distribution of each of the four primary outcomes by treatment. Valuation (panel~a) is bimodal in every arm, with a large share of participants at the low end (\$0-\$5) and at the top of the scale (\$50 or above). In the \textit{Chat} treatment, however, a greater share of participants have the lowest valuation and fewer have the highest valuation. Mean valuations are \$27.93 in \textit{Filter}, \$24.97 in \textit{Chat}, and \$30.44 in \textit{Hybrid}. \textit{Chat} valuations are significantly lower than \textit{Hybrid} (rank sum test, $p = 0.024$), while \textit{Filter} is statistically indistinguishable from either treatment ($p = 0.232$ vs.\ \textit{Chat}, $p = 0.316$ vs.\ \textit{Hybrid}). Search time (panel~b) is right-skewed everywhere, but the \textit{Chat} distribution has a visibly heavier right tail suggesting that a subset of participants continue iterating with the LLM for a long time to find an event they like. \textit{Chat} subjects spend $3.57$ minutes on average, which is about $30\%$ longer than \textit{Filter} ($2.75$ min; $p < 0.001$) and about $22\%$ longer than \textit{Hybrid} ($2.91$ min; $p < 0.001$), while \textit{Filter} and \textit{Hybrid} do not differ significantly ($p = 0.836$). 

We observe similar results for satisfaction scores. For satisfaction with the chosen event (panel~c), the \textit{Chat} distribution is shifted towards lower ratings. \textit{Chat} subjects rate their selected event significantly lower than \textit{Filter} ($5.32$ vs.\ $5.88$; $p = 0.014$) and lower still than \textit{Hybrid} ($6.04$; $p < 0.001$), while \textit{Filter} and \textit{Hybrid} are statistically indistinguishable ($p = 0.315$). Satisfaction scores for the process (panel~d) are similar. In sum, pure \textit{Chat} appears to be the worst-performing arm, while the differences between \textit{Filter} and \textit{Hybrid} are minimal on every measure. This provides some initial evidence that layering a conversational interface on top of the structured filter does not improve outcomes relative to filters alone, but it also does no harm. In contrast, a conversational interface alone leads to lower valuations, longer search times, and lower satisfaction with both the outcome and the experience. We next present formal tests of our hypotheses.

\subsection{Regression Results}\label{sec:res-regression}
We regress each of the four main outcome variables (valuation, outcome satisfaction, search time, and process satisfaction) on the treatment and a vector of covariates (age, gender, education, income, travel distance to NYC, and NY/NJ residence) and session and experimental wave controls. The results are in Table~\ref{tab:reg_main}.

Column~(1) shows that valuation in neither \textit{Chat} nor \textit{Hybrid} differs from \textit{Filter} at conventional significance levels, but \textit{Hybrid} produces higher valuation than \textit{Chat}. Column~(2) shows that \textit{Chat} participants are significantly less satisfied with their chosen event than both \textit{Filter} and \textit{Hybrid} participants, who are indistinguishable from each other. Column~(3) indicates that users in the \textit{Chat} treatment take about $62$ seconds longer than \textit{Filter}, while \textit{Hybrid} is statistically indistinguishable from \textit{Filter}. Column~(4) shows that \textit{Chat} participants are significantly less satisfied with the search process than both \textit{Hybrid} and \textit{Filter}, which are again indistinguishable. In sum, \textit{Hybrid} significantly outperforms \textit{Chat} on all outcomes except search time and \textit{Filter} significantly outperforms \textit{Chat} on all outcomes except valuation. Furthermore, there are no significant differences between \textit{Filter} and \textit{Hybrid}.

\begin{table}[bt]
\centering
\caption{Treatment Effects}\label{tab:reg_main}
\small
\sbox{\regtablebox}{%
\begin{tabular}{l *{4}{>{\centering\arraybackslash}m{2.5cm}}}
\toprule
\multicolumn{1}{r}{Dep.\ var.:} & \textit{Valuation} & \textit{Outcome Satisfaction} & \textit{Search Time} & \textit{Process Satisfaction} \\
\cmidrule(lr){2-2}\cmidrule(lr){3-3}\cmidrule(lr){4-4}\cmidrule(lr){5-5}
\multicolumn{1}{r}{Estimator:} & Interval & OLS & OLS & OLS \\
 & (1) & (2) & (3) & (4) \\
\midrule
\textit{Filter} Treatment & Omitted & Omitted & Omitted & Omitted \\
\addlinespace
\textit{Chat} Treatment & $-$4.203 & $-$0.635$^{***}$ & 61.914$^{***}$ & $-$0.699$^{***}$ \\
 & (3.576) & (0.191) & (23.216) & (0.200) \\
\addlinespace
\textit{Hybrid} Treatment & 3.518 & 0.182 & 21.200 & $-$0.230 \\
 & (3.875) & (0.173) & (26.741) & (0.198) \\
 \midrule
\textit{Hybrid} Treatment $-$ \textit{Chat} Treatment & 7.721$^{**}$ & 0.817$^{***}$ & $-$40.714 & 0.469$^{**}$ \\
 & (3.527) & (0.185) & (24.865) & (0.206) \\
\midrule
Demographic controls & Yes & Yes & Yes & Yes \\
Session FE & Yes & Yes & Yes & Yes \\
Experimental Wave FE & Yes & Yes & Yes & Yes \\
Observations & 362 & 362 & 362 & 362 \\
$R^2$ (adj.) & --- & 0.069 & 0.032 & 0.053 \\
\bottomrule
\end{tabular}}%
\usebox{\regtablebox}\par
\vspace{3pt}
\begin{minipage}{\wd\regtablebox}
\footnotesize\linespread{0.9}\selectfont
\textit{Notes:} Standard errors in parentheses (HC3 robust for OLS). Column~1 is an interval regression. Columns 2--4 are OLS. Demographic controls are age, gender, income, education, NYC proximity, and NY/NJ residence. $^{***}p<0.01$, $^{**}p<0.05$, $^{*}p<0.1$.\par
\end{minipage}
\end{table}

Since our only incentivized outcome measure is valuation, it is worth examining the treatment effect on this measure more closely. In particular, recall that in Table~\ref{tab:reg_main}, col. (1) we used an interval regression.\footnote{We do not report OLS and other censored variable regression (e.g., Tobit) estimates for brevity, but note that they are similar in magnitude and have the same significance levels as the estimates reported in the main text.} The interval specification is well-suited for our multiple-price list elicitation approach given that we do not observe precise valuation, but instead observe the smallest monetary amount that would prompt a decision-maker to accept money over a ticket to their chosen event. Interval regression is the standard approach for this elicitation format \citep{CameronHuppert1989, HarrisonRutstrom2008}. Furthermore, given the censoring of the dependent variable at the left and the right tail of the distribution, it is informative to examine the latent treatment effects (the effects that would be observed in the absence of censoring), as well as examine the treatment effects on the probability of censoring. 

\begin{figure}[tbp]
\caption{Post-Estimation Results (All Based on Table~\ref{tab:reg_main}, Col. 1)}
\label{fig:reg_ame}
\centering
\includegraphics[width=\linewidth]{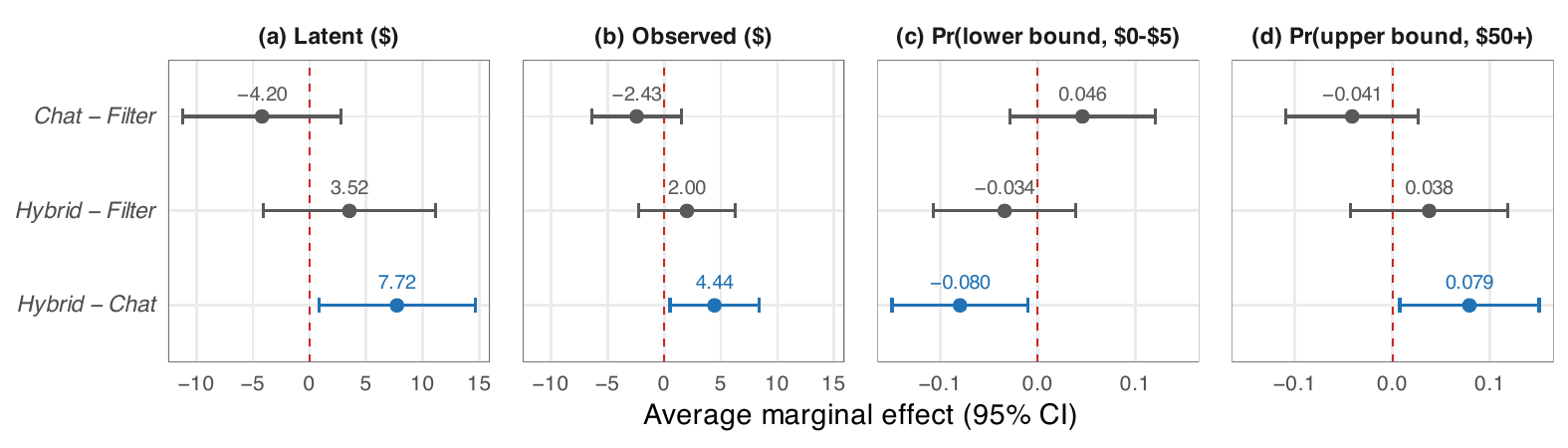}
\end{figure}

Panels (a) and (b) of Figure~\ref{fig:reg_ame} report the latent effects and the observed effects; panels (c) and (d) report the treatment effects on the probability of being censored (i.e., being at the lower or at the upper bound of the valuation scale). Consistent with Table~\ref{tab:reg_main}, only the contrast between \textit{Hybrid} and \textit{Chat} is significantly different from zero. In particular, \textit{Hybrid} increases the latent valuation by \$7.72 and the observed valuation by \$4.44 (the observed effect is smaller because a large proportion of valuations are at the upper bound).  Panel (c) of Figure~\ref{fig:reg_ame} shows that the probability of minimal valuation (\$0-\$5) is $8.0$ percentage points higher in \textit{Chat} than in \textit{Hybrid}. Vice versa, the probability of being above \$50 is $7.9$ percentage points lower in \textit{Chat} than in \textit{Hybrid}. That is, the \textit{Hybrid} interface moves the entire distribution of valuations including the lower bound, the interior and the upper bound to the right. In contrast, neither \textit{Chat} nor \textit{Hybrid} differs from \textit{Filter} on any margin.

\smallskip
\noindent\textbf{Result 1:} H1 is partially supported. Consistent with H1, \textit{Hybrid} valuations and outcome satisfaction are higher than in \textit{Chat}. Contrary to H1, \textit{Chat} does not outperform \textit{Filter}. Furthermore, \textit{Filter} performs similarly to \textit{Hybrid}.
\smallskip

\smallskip
\noindent\textbf{Result 2:} H2 is partially supported. Consistent with H2, \textit{Chat} requires the most search time and yields the lowest process satisfaction. Contrary to H2, \textit{Filter} and \textit{Hybrid} are statistically indistinguishable on both measures.
\smallskip

In addition to making treatment comparisons, we briefly report which search modalities participants use when more than one is available (i.e., in the \textit{Filter} and \textit{Hybrid} conditions) and how the use of these modalities is related to valuations. In \textit{Filter}, 45\% of participants used only filters, 52\% combined filters with text entry, and 3\% used only text-entry. The corresponding shares in \textit{Hybrid} are 51\%, 45\%, and 4\%. Valuation differences are minimal between the two large subgroups within \textit{Filter} (\$25.9 for filters only versus \$29.6 for both, rank sum $p = 0.30$), and within \textit{Hybrid} (\$30.9 versus \$29.2, $p = 0.74$). This is also true for each subgroup comparison across the two arms: participants who only used filters have similar valuations under \textit{Filter} and \textit{Hybrid} (\$25.9 versus \$30.9, $p = 0.20$), and also for those who combined filters with text entry (\$29.6 versus \$29.2, $p = 0.96$). 

\subsection{Model Performance and Search Intent}\label{sec:res-model-intent-model}
We next address Research Question \textit{(iii)}. In particular, we test (1) whether the treatment effects depend on which AI model powers the chat and (2) whether the treatment effects depend on the user's search intent. 

\subsubsection{AI Model Performance}\label{sec:res-model-performance}
Our first analysis concerns model performance. In particular, recall that within the \textit{Chat} treatment arm, the AI model was randomized: each participant was randomly assigned either to an older, legacy model (\texttt{gpt-4.1-mini}, released in 4/2025) or to one of two frontier models (\texttt{gpt-5.2}, released in 12/2025, and \texttt{gpt-5.4}, released in 3/2026). Recall that the router and retrieval stages were identical across models (\textsection\ref{sec:meth-rag}), so any difference in outcomes would arise from the text generation step alone.

\begin{figure}[b]
\caption{Pairwise AI-model contrasts (within the \textit{Chat} treatment).}
\label{fig:model_comparison}
\centering
\includegraphics[width=\linewidth]{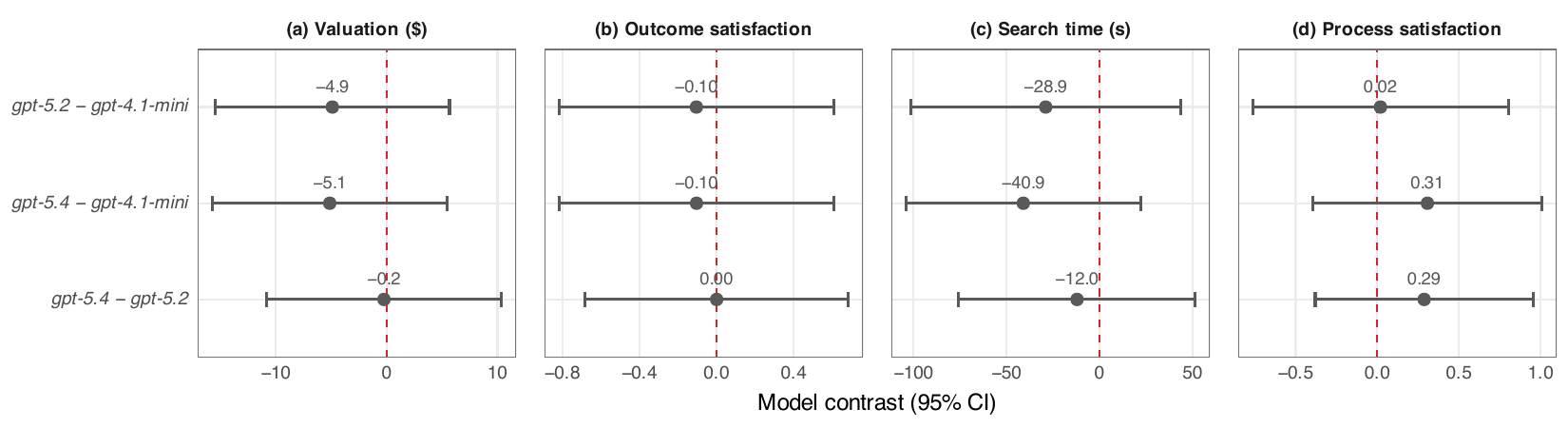}
\end{figure}

Figure~\ref{fig:model_comparison} reports the pairwise contrasts between the three models on each primary outcome, estimated from the regressions in Table~\ref{tab:model_reg} ($95\%$ confidence intervals). The three models are statistically indistinguishable on every outcome: every interval includes zero. Search time is closest to being significantly different between models, with \texttt{gpt-4.1-mini} being the slowest. Note, however, that this model also leads to directionally higher valuations than the more recent models; see panel (a). Overall, there is no evidence that a more capable model improves search performance or reduces search costs in our setting. 

\subsubsection{Search Intent}\label{sec:res-model-intent-intent}
Our second analysis concerns the role of search intent. In particular, in the exit survey, participants reported whether they approached the task with no particular event in mind (\emph{open-ended}, $35\%$ of the sample), somewhere in between (\emph{in-between}, $47\%$), or with a specific event in mind (\emph{specific}, $18\%$). We add the treatment interactions with this three-level intent measure to the regression specification in Table~\ref{tab:reg_main}. We focus on valuation as our outcome measure, but results are broadly consistent across all our dependent variables. 

Figure~\ref{fig:intent_margins} plots the three pairwise treatment contrasts on valuation at each intent level (See Appendix \textsection\ref{app:intent} Table~\ref{tab:intent_appx} for the regression table). The figure shows that the treatment differences are found only among open-ended searchers, for whom \textit{Chat} is \$15.89 below \textit{Filter} and \$21.14 below \textit{Hybrid} (both $p<0.01$). In contrast, for in-between and specific searchers, treatment effects are not statistically different from zero. That is, the conversational interface hurts the consumers who want to explore, rather than those who are trying to retrieve a specific event.

\smallskip

\noindent\textbf{Result 3:} The effects of conversational AI do not depend on the AI model (\texttt{gpt-4.1-mini}, \texttt{gpt-5.2}, or \texttt{gpt-5.4}). \textit{Chat} lowers valuations among users performing open-ended exploration, whereas users retrieving a specific target do just as well under \textit{Chat} as in the other conditions.
\smallskip

\begin{figure}[b]
\caption{Treatment Effects on Valuation by Search Intent}
\label{fig:intent_margins}
\centering
\includegraphics[width=\linewidth]{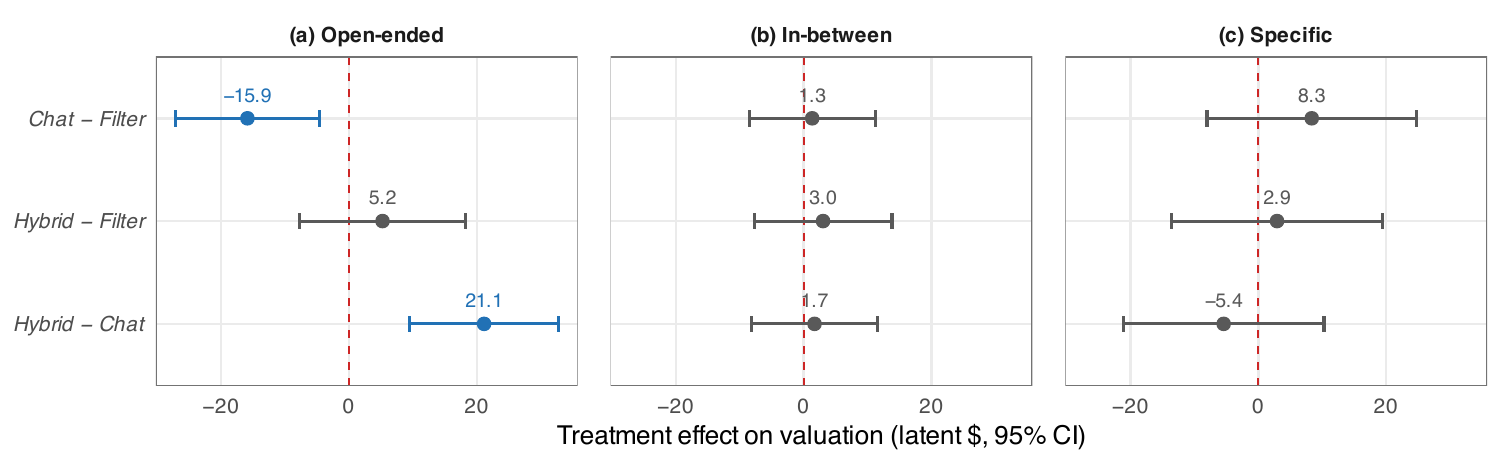}
\end{figure}

\section{Mechanisms}
\label{sec:Mechanisms}
We document two mechanisms that help explain behavior and performance in the \textit{Chat} treatment. The first mechanism builds on the idea of an \textit{echo chamber}, popularized in the research on news and social media consumption \citep{cinelli2021echo}. Consistent with this idea, we show that the \textit{Chat} interface does not provide the user with sufficient exposure to the catalog. The second mechanism, which we will refer to as \textit{prompting fatigue}, builds on the idea that, to be useful, generative AI requires the user to spend an appropriate amount of effort and attention on inspecting the output of the AI model \citep{DellAcquaEtAl2023,SpathariotiEtAl2025}. We show that the \textit{Chat} interface reallocates effort and attention away from inspecting results and towards composing queries and waiting for responses, and that underinvestment of effort leads to poor search outcomes.{\footnote{In \textsection\ref{app:process-full}, we present a detailed list of process variables and their correlations with valuation.}}

\subsection{\textit{Chat} Creates an \textit{Echo Chamber}: It Reduces Search Breadth and Increases Out-of-Catalog Queries}\label{sec:mech-echo}
The \textit{echo chamber} metaphor is most often used in the context of news and social media, where personalized sorting and recommendations have been shown to reinforce the beliefs held by the user rather than expose them to new, sometimes conflicting opinions or ideas \citep{cinelli2021echo}. More recently, this concept has also been used to describe interactions with Generative AI, mainly for queries related to political content \citep{sharma2024}. Different from social media, our platform does not have user profiles, nor does it attempt to make an inference about user intent. Instead, it is programmed to simply return the list of events with the best match to the users' queries (as well as accompanying text). Nonetheless, an \textit{echo chamber} can still arise in our setting. This is because users compose queries based on what they believe is contained in the catalog, and retrieval reflects those beliefs back as search results. The AI output can therefore limit what the user sees by showing them a narrow slice of the catalog that includes the types of items the user already had in mind. Furthermore, because the user is not sufficiently familiar with the contents of the catalog, their queries may land outside the catalog altogether.

To examine the \textit{echo chamber} effect in our data we focus on two outcomes: search breadth and out-of-catalog search. We define search breadth as the number of distinct segments (among \textit{Music}, \textit{Sports}, \textit{Arts \& Theater}) that a user searches within a session. We define out-of-catalog search as the number of instances in which a user searches for an event, venue, artist, sports team, genre, subgenre, or segment that does not appear in the list of events scraped from the platform. We use the interaction logs collected from the experiment to construct these variables, and present a detailed description of their construction in \textsection\ref{app:mech_variables}. 

\begin{figure}[bt]
\caption{\textit{Echo Chamber} Mechanism}
\label{fig:echo_wtp}
\centering
\includegraphics[width=\linewidth]{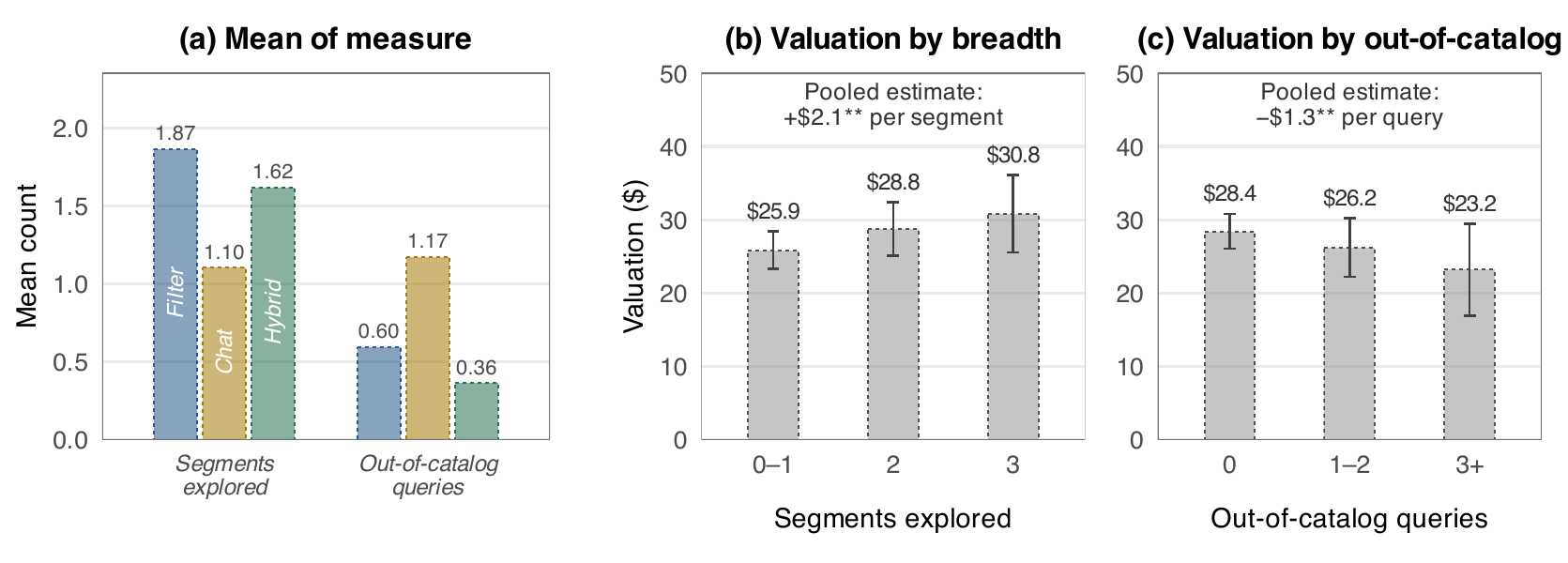}

\vspace{3pt}
\begin{minipage}{0.99\linewidth}
\footnotesize\linespread{0.9}\selectfont \emph{Note.} Panel (a) shows the mean of each search measure by treatment arm. Panel (b) shows the effects of the number of segments explored on valuation. Panel (c) shows the effects of the number of out-of-catalog queries on valuation. Estimates and their significance levels are based on average marginal effects from pooled interval regressions in Table~\ref{tab:echo_spec}. $^{*}\,p<0.1$, $^{**}\,p<0.05$, $^{***}\,p<0.01$.
\end{minipage}
\end{figure}

Figure~\ref{fig:echo_wtp} summarizes the analysis: panel (a) shows the mean for each measure by treatment arm, and panels (b) and (c) show mean valuation by binned level of each measure, together with the pooled valuation effects based on interval-regression specifications (Appendix Table~\ref{tab:echo_spec} reports the regression estimates). Panel (a) shows that participants in the \textit{Chat} arm score differently from the other two treatments on both process measures. \textit{Chat} participants search 1.10 distinct segments on average, compared with 1.87 in \textit{Filter} and 1.62 in \textit{Hybrid} (rank sum tests of \textit{Chat} against each of the other arms, both $p < 0.001$). Conversely, if we examine out-of-catalog search, \textit{Chat} participants submit 1.17 out-of-catalog queries on average, compared with 0.60 in \textit{Filter} and 0.36 in \textit{Hybrid} (rank sum tests of \textit{Chat} against each arm, $p = 0.002$ and $p < 0.001$). These differences suggest that the AI interface exposes users to a narrower slice of the catalog, and more of their queries concern events (or event types) that are not available in the catalog. 

Panels (b) and (c) tie these process measures to valuations. Valuations increase monotonically with search breadth, from \$25.9 among participants who explore at most one segment to \$28.8 at two segments and \$30.8 at three. Conversely, valuations decline with out-of-catalog search, from \$28.4 among participants with no out-of-catalog queries to \$26.2 for one or two such queries and \$23.2 for three or more. Exploring each additional segment is associated with a \$2.1 higher valuation ($p = 0.047$) and each additional out-of-catalog query with a \$1.3 lower valuation ($p = 0.028$) when pooled across treatments. See Table~\ref{tab:echo_spec} for estimation details.

\subsection{\textit{Chat} Leads to \textit{Prompting Fatigue}: It Shifts Time from Inspecting Results to Composing Queries}\label{sec:mech-verif}

We next examine our second mechanism, which we refer to as \textit{prompting fatigue}. In particular, we show that the search interface can affect search outcomes by changing how users allocate their effort and attention across different activities of the search process, i.e., composing queries, waiting for AI responses, and inspecting results. 
Prompting fatigue builds on the idea that attention is depletable and deteriorates as a task is prolonged \citep{mackworth1948breakdown}. In our setting the depletion happens because the chat interface prolongs the search by requiring users to spend more time on composing and waiting; users therefore reach the results later and more depleted. As a result, they are more likely to select one of the first events displayed, which is in turn associated with lower valuations.

\begin{figure}[tb] 
\centering
\caption{\textit{Prompting Fatigue} Mechanism}
\label{fig:attention_panels}
\includegraphics[width=\textwidth]{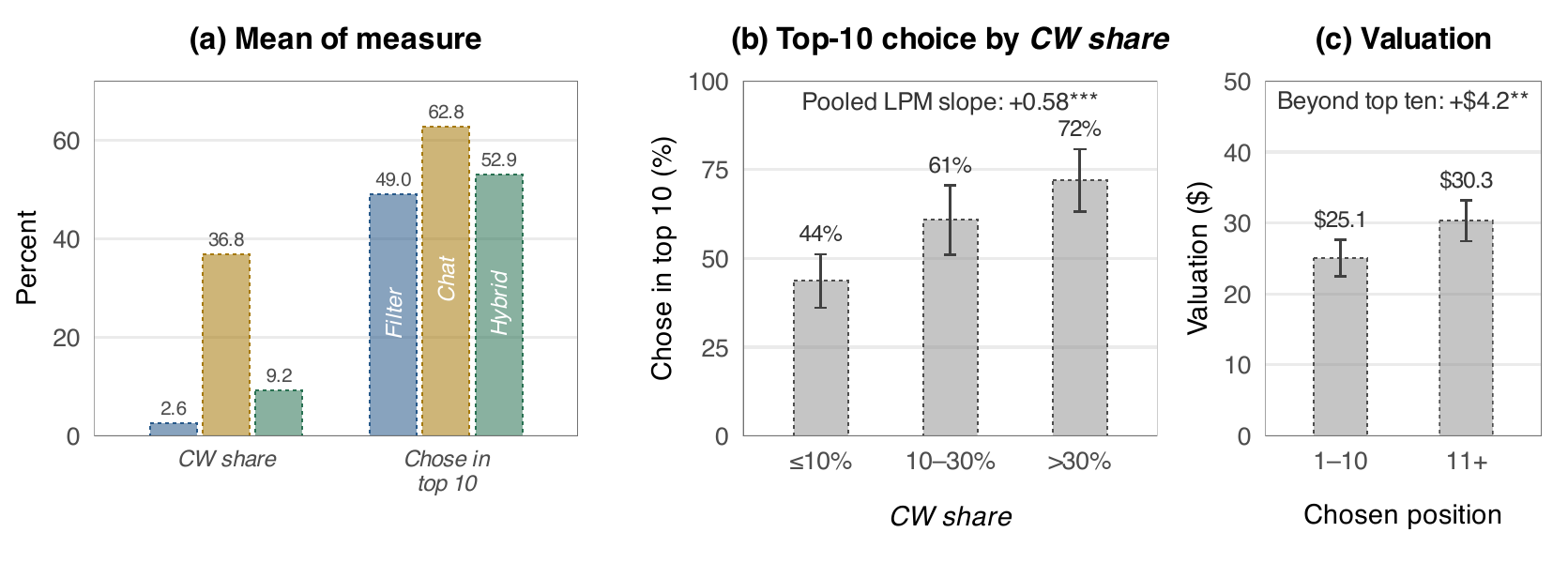}
\vspace{3pt}
\begin{minipage}{0.99\linewidth}
\footnotesize\linespread{0.9}\selectfont \emph{Note.} Panel (a) shows mean \textit{CW share} (the share of the session spent composing queries and waiting) and the share of participants choosing one of the top ten displayed events. Panel (b) shows the effect of \textit{CW share} on choosing from the top ten. Panel (c) shows mean valuation by the position of the chosen event in the displayed list (error bars $95\%$ CI). The annotated effects are based on linear-probability slopes in panel (b) and the pooled average marginal effect in panel (c), both estimated with the demographic controls, pooled across arms (see \textsection\ref{app:attention} for complete regression results). 
$^{*}\,p<0.1$, $^{**}\,p<0.05$, $^{***}\,p<0.01$. 
\end{minipage}
\end{figure}

Figure~\ref{fig:attention_panels}  presents the evidence supporting this mechanism. For simplicity, we split each session into two types of activities: composing queries and waiting for AI responses (CW), and inspecting the returned results. The \textit{CW share} is the fraction of the session spent composing queries and waiting rather than examining results. We then examine how \textit{CW share} affects the extent to which a participant chooses one of the first few events displayed to them.\footnote{We focus on the \emph{share} of time spent on different activities (as opposed to absolute durations) because our primary interest is in how the interface affects the allocation of a user's attention across activities. Absolute durations can reflect both the differences caused by the interface and variation in users' overall time budgets, while shares normalize the time budgets. Formally, let $t_c$, $t_w$, and $t_i$ denote the time spent composing queries, waiting for responses, and inspecting results, which together account for the session; the \textit{CW share} is $s = (t_c + t_w)/(t_c + t_w + t_i)$. Composing time $t_c$ is approximated as typed characters divided by a typing rate of three characters per second, waiting time $t_w$ is the measured response time of the AI, and inspecting time $t_i$ is the residual of the session. In \textit{Filter}, $t_w = 0$ mechanically, since a drop-down returns results instantly.} Figure~\ref{fig:attention_panels}(a) reports the mean \textit{CW share} and the share of participants whose final choice is one of the first 10 displayed events, by treatment arm.\footnote{The specific choice of 10 events is based on pagination in the experiment -- recall that up to 10 events can be displayed on each page of returned results and to continue to the next 10 events the user must click on the ``next 10 events'' button. However, the results are robust to an alternative specification where we focus on the top three events instead of the top 10.} Notably, in \textit{Filter}, composing and waiting takes up only $2.6\%$ of the session, compared with $36.8\%$ in \textit{Chat} (rank sum tests of \textit{Chat} against each arm, both $p < 0.001$), with \textit{Hybrid} being in between. The treatments also differ in how often participants choose from the top of the list. In particular, the right side of panel (a) shows that the probability of choosing an event in a top ten position is $49.0\%$ in \textit{Filter} vs. $62.8\%$ in \textit{Chat}, with \textit{Hybrid} at $52.9\%$ (rank sum tests of \textit{Chat} against \textit{Filter} and \textit{Hybrid}, $p = 0.028$ and $p = 0.116$). 

Panel (b) shows that the share of participants choosing from the top ten increases from $44\%$ among those with \textit{CW shares} below $10\%$, to $61\%$ for \textit{CW shares} between $10\%$ and $30\%$, and to $72\%$ above $30\%$. Pooled across arms, the probability of choosing one of the first 10 events increases by $0.58$ per unit of \textit{CW share} ($p < 0.001$). Furthermore, Panel (c) shows that choosing one of the first few events displayed reduces valuations with the average estimated gap of \$4.2 ($p = 0.013$).

While our analysis of \textit{prompting fatigue} focuses on relative time allocation, examining the absolute times spent on each activity helps better understand this mechanism. In particular, each additional minute of composing and waiting increases the probability of the chosen event being in the top 10 displayed events ($+0.10$ within \textit{Chat}, $p = 0.014$), while the time spent inspecting has no significant effect ($p = 0.46$ within \textit{Chat}). This suggests that the main driver is the effort spent \textit{before} the results are reached, rather than a shortage of time spent inspecting them. In other words, participants who accumulate composing and waiting time reach the results with depleted patience and choose one of the first events shown. Moreover, both components contribute to the fatigue, as both the composing share ($+0.76$, $p = 0.007$) and the waiting share ($+0.58$, $p = 0.044$) predict a top ten choice about equally within \textit{Chat}. In other words, the mechanism is driven not only by passively waiting for AI to generate results but also by the effort required to build queries.

\subsection{Discussion}
The two discussed mechanisms provide plausible explanations for the observed treatment effects, but each raises questions about what the data does and does not show.

First, the effects of breadth and out-of-catalog queries on valuations should be interpreted with care. This is because the treatments differ in the cost of learning about the catalog. Under \textit{Filter} and \textit{Hybrid}, the drop-down lists the event categories offered by the platform, and moving between them takes a single click. In contrast, under \textit{Chat}, to explore a new category one needs to produce a new query (i.e., pivot away from the current conversation). Breadth therefore does not mean the same across treatment arms as it can be achieved nearly for free in one case but requires costly effort in the other. We thus read the treatment effects on the process measures (breadth, out-of-catalog searches) as causal, since they are caused by random treatment assignment, whereas the link between these measures and valuations may be non-causal as it combines the returns to exploration with differences in effort and engagement across users. For example, our data cannot tell whether exogenously increasing a \textit{Chat} user's breadth would improve this user's valuation. The interface does determine, however, how a user can respond once a query lands outside the catalog. The menu lets \textit{Filter} and \textit{Hybrid} users redirect their search immediately, while \textit{Chat} users are likely to continue querying variations of their initial target. Indeed, visibility into the catalog makes an out-of-catalog query less likely in the first place. For example, one user in our sample searched for ``\textit{food related events}'' in \textit{Chat}; had the drop-down shown that the catalog contains only Music, Sports, and Arts \& Theater, this search would likely not have occurred. 

Our second mechanism (\textit{prompting fatigue}) may, at first glance, appear counterintuitive. Conventionally, a ranked list displays the best matches first, and participants who choose from the top ten should do at least as well as those who search deeper. Unlike traditional recommendation systems where the rank reflects the relative value of the alternative \citep{allu2025ranking}, RAG-based interfaces such as ours order results by similarity to the query (see \textsection\ref{sec:meth-rag} and \textsection\ref{app:rag-embedding}). As a result, the position of an event in the retrieved list does not always provide much information about its relative value to the user. This is particularly true for queries that are more exploratory. For example, for a query like \emph{``something fun this weekend''}, the first ten positions are, in value terms, an arbitrary subset of the results. Therefore, for these types of queries, the larger the set of events a user considers, the higher the chance of finding a good match. To test this logic, in \textsection\ref{app:attention-specificity} we split the sample by query specificity (did the participant ever name a specific event, artist, or team?) and find that the valuation gap is only observed among exploratory searchers ($+\$6.7$ for choosing beyond the top ten, $p = 0.001$) while being insignificant among specific searchers ($+\$2.1$, $p = 0.48$). \textit{Prompting fatigue} is therefore particularly harmful for exploration. 

Just as with our first mechanism, \textit{prompting fatigue} should be interpreted with care. The effect of treatment on \textit{CW share} is causal as it results from random assignment, whereas the effects of \textit{CW share} on choosing a top ten event, and from that choice to lower valuations, need not be. In particular, our data cannot tell a participant who is fatigued and simply picks one of the first displayed results from a participant who rationally spends minimal effort on inspecting the results because the retrieved list contains nothing better beyond the first few positions. In the latter case, choosing from the top may be the best strategy, and the low valuation is simply due to the retrieved event list rather than due to how attention was allocated.

Finally, while the two mechanisms (\textit{echo chamber} and \textit{prompting fatigue}) are based on different measurements, we believe that they share a common root. With a menu-based interface, the user needs only a few clicks to narrow down a choice set from a large catalog of heterogeneous events. But, with a conversational interface, the user must produce the choice set by describing what they want and wait for the results. This strains the consumer's limited attention in two ways. First, in terms of the content seen, the choice sets can only be as good as the queries that generate them, which are composed from what the user already has in mind; this is the \textit{echo chamber} mechanism. Second, in terms of the search process, the need to produce results absorbs attention that would otherwise be spent examining them; this is the \textit{prompting fatigue} mechanism. Both mechanisms are, in other words, costs of making the user become the producer of their own search.

\section{Practical Implications and Outlook}\label{sec:Discussion}
Our experiment compared three versions of an event-discovery platform: a traditional interface with filters and keyword search, a conversational AI interface, and a hybrid offering both. Our results help explain the differences in AI adoption observed across platforms in practice and offer prescriptive guidance for those still deciding, which we discuss in \textsection\ref{subsec:implications}. Moreover, our results point to several open questions, which we discuss in \textsection\ref{subsec:outlook}.

\subsection{Practical Implications}\label{subsec:implications}

Our comparison of the three interfaces was motivated by the rapid deployment of conversational search across digital platforms, and, just as in our experiment, the value of these tools in practice depends on how they are implemented and in what settings they are used. Indeed, as some platforms have embraced conversational search (as described in \textsection\ref{sec:Intro}), others have been more hesitant to move forward with it. For example, Airbnb decided against conversational search, with the reasoning that structured filters are better suited to help users explore their catalog than free-form chat.\footnote{Airbnb rejects AI chatbot: \url{https://skift.com/2024/11/21/airbnbs-ai-plan-no-chatbot-more-personalization/}.} Similarly, Netflix has piloted conversational search for over a year without a broad rollout, citing limited customer comfort with the modality.\footnote{See \url{https://about.netflix.com/en/news/unveiling-our-innovative-new-tv-experience} for the initial announcement, and \url{https://toomuchtv.substack.com/p/too-much-tv-netflixs-effort-to-improve} for commentary on its adoption.}

We believe that the divide between platforms that have successfully deployed AI tools and those that have not is not coincidental. Rather, it is consistent with the moderating effects of search intent identified in our experiment. Travel booking platforms that have moved quickly tend to serve users who have a specific target, such as booking a hotel in a specific location or finding a flight for a specific route. These use cases are defined by a well-defined, pre-conceived target -- precisely the scenarios where we find conversational AI to do little harm. In contrast, the platforms that have been hesitant to deploy AI serve users who are more interested in open-ended exploration. Both Netflix and Airbnb sell experiences, and experiences are found by browsing, which is exactly the type of search where chat appears to perform worst. 

In addition to providing a plausible explanation for some of the AI adoption patterns observed in practice, our results can be used to make several prescriptive recommendations. In particular, our results suggest that the relevant question for platforms is not \textit{whether} but \textit{how} to deploy the chat modality. Layering a conversational interface on top of a structured filter, as in our \textit{Hybrid} condition, has no detectable cost. In contrast, replacing the filter with chat lengthens search, narrows the portion of the catalog users see, and lowers the value of what they choose. Importantly, platforms need not invest in state-of-the-art LLMs to offer this functionality. Our results show that valuations are invariant to the model powering the chat (\texttt{gpt-4.1-mini}, \texttt{gpt-5.2}, \texttt{gpt-5.4}). In other words, users are not sensitive to the specific language used by the AI to describe its search results. This suggests that retrieval quality (as opposed to response generation) is what drives success or failure in the deployment of these AI systems, and even a lightweight AI model, such as \texttt{gpt-4.1-mini}, can perform RAG-based retrieval effectively.

As a final, methodological takeaway, we show that incentivized elicitation methods (such as the BDM method, popularized in behavioral science) and survey-based satisfaction scores agree on most measures when it comes to evaluating AI search performance. Platforms (for whom satisfaction surveys are standard practice while incentive-compatible elicitation is not) can thus evaluate the costs and benefits of deploying AI tools from survey data alone.

\subsection{Outlook}\label{subsec:outlook}
Our results suggest several potential follow-up questions. First, the design space between a bare chat window and a structured menu is largely unexplored. Our mechanisms point to plausible candidates for future study, such as displaying the catalog's structure inside the conversation, reducing the time cost of composing and waiting (e.g., through prompt suggestions), or routing queries to the modality that suits them best. Second, our participants used the interfaces once. Regular users may learn to compose better queries and use AI recommendations more effectively. Finally, search interfaces may perform differently depending on the device, and our study only focused on desktop devices. A mobile device with a smaller screen may be less useful for menu-based search and more suited to chat-based interactions.

The RAG architecture used in our experiment also offers a more general template for studying AI-assisted decision-making in both settings traditionally examined by OM scholars and emerging settings motivated by recent developments in practice. An example of the former is the problem of a production planner who makes inventory ordering decisions \citep[e.g.,][]{SchweitzerCachon2000,BoltonKatok2008}  or forecasts demand based on past sales \citep{KremerMoritzSiemsen2011,MoritzSiemsenKremer2014}. An important (but understudied) step in this process is choosing which past products and periods are sufficiently comparable to serve as a reference class for the forecast. Prior work has shown that managers make errors in choosing such data \citep{DiSorboFerreiraBalakrishnanTong2025}. Conversational AI tools can make this recommendation to managers in natural language, which raises the question of whether such assistance improves the quality of forecasts. In emerging settings, firms are increasingly giving employees access to conversational AI tools that can be used to query internal data, often via RAG-based protocols.\footnote{For example, MicroStrategy, a business intelligence software provider, develop RAG-based AI tools that allow customers to translate firm data and performance metrics into natural-language responses. See \url{https://www.strategy.com/press/microstrategy-releases-auto-the-customizable-ai-bot-that-makes-enterprise-analytics-accessible-to-anyone-building-on-microstrategy-ai-03-26-2024}.} Whether such tools improve the decisions they support, and how their interfaces should be designed, is an open question. RAG-based experimental designs provide a way to study both these questions where an experimenter can control the output of the AI tool (unlike a standard LLM which can produce stochastic outputs even for the same prompt). RAG-based designs can thus help advance future behavioral work by improving experimental control without sacrificing realism.

\bibliographystyle{informs2014} 
\bibliography{references} 

\begin{thebibliography}{82}
\providecommand{\natexlab}[1]{#1}
\providecommand{\url}[1]{\texttt{#1}}
\providecommand{\urlprefix}{URL }

\bibitem[{Adam et~al.(2021)Adam, Wessel, \protect\BIBand{} Benlian}]{AdamWesselBenlian2021}
Adam M, Wessel M, Benlian A (2021) {AI}-based chatbots in customer service and their effects on user compliance. \emph{Electronic Markets} 31(2):427--445.

\bibitem[{Allu \protect\BIBand{} Gaur(2025)}]{allu2025ranking}
Allu R, Gaur V (2025) Ranking quality and user engagement on an online b2b platform. \emph{Available at SSRN 4684782} .

\bibitem[{Andersen et~al.(2006)Andersen, Harrison, Lau, \protect\BIBand{} Rutstr\"om}]{AndersenHarrisonLauRutstrom2006}
Andersen S, Harrison GW, Lau MI, Rutstr\"om EE (2006) Elicitation using multiple price list formats. \emph{Experimental Economics} 9(4):383--405.

\bibitem[{Antonides et~al.(2002)Antonides, Verhoef, \protect\BIBand{} van Aalst}]{AntonidesVerhoefVanAalst2002}
Antonides G, Verhoef PC, van Aalst M (2002) Consumer perception and evaluation of waiting time: A field experiment. \emph{Journal of Consumer Psychology} 12(3):193--202.

\bibitem[{{AWS}(2024)}]{AWS2024RAG}
{AWS} (2024) What is {RAG (Retrieval-Augmented Generation)}? AWS Machine Learning Documentation.

\bibitem[{Bakos(1997)}]{Bakos1997}
Bakos JY (1997) Reducing buyer search costs: Implications for electronic marketplaces. \emph{Management Science} 43(12):1676--1692.

\bibitem[{Balakrishnan et~al.(2026)Balakrishnan, Ferreira, \protect\BIBand{} Tong}]{BalakrishnanFerreiraTong2026}
Balakrishnan M, Ferreira KJ, Tong J (2026) Human-algorithm collaboration with private information: Na{\"i}ve advice-weighting behavior and mitigation. \emph{Management Science} 72(1):265--284.

\bibitem[{Batra \protect\BIBand{} Ahtola(1990)}]{BatraAhtola1990}
Batra R, Ahtola OT (1990) Measuring the hedonic and utilitarian sources of consumer attitudes. \emph{Marketing Letters} 2(2):159--170.

\bibitem[{Becker et~al.(1964)Becker, DeGroot, \protect\BIBand{} Marschak}]{BeckerDeGrootMarschak1964}
Becker GM, DeGroot MH, Marschak J (1964) Measuring utility by a single-response sequential method. \emph{Behavioral Science} 9(3):226--232.

\bibitem[{Becker(1965)}]{Becker1965}
Becker GS (1965) A theory of the allocation of time. \emph{The Economic Journal} 75(299):493--517.

\bibitem[{Bettman et~al.(1998)Bettman, Luce, \protect\BIBand{} Payne}]{BettmanLucePayne1998}
Bettman JR, Luce MF, Payne JW (1998) Constructive consumer choice processes. \emph{Journal of Consumer Research} 25(3):187--217.

\bibitem[{Binz \protect\BIBand{} Schulz(2023)}]{BinzSchulz2023}
Binz M, Schulz E (2023) Using cognitive psychology to understand {GPT-3}. \emph{Proceedings of the National Academy of Sciences} 120(6):e2218523120.

\bibitem[{Bolton \protect\BIBand{} Katok(2008)}]{BoltonKatok2008}
Bolton GE, Katok E (2008) Learning by doing in the newsvendor problem: A laboratory investigation of the role of experience and feedback. \emph{Manufacturing \& Service Operations Management} 10(3):519--538.

\bibitem[{Brand et~al.(2023)Brand, Israeli, \protect\BIBand{} Ngwe}]{BrandIsraeliNgwe2023}
Brand J, Israeli A, Ngwe D (2023) Using {LLM}s for market research Harvard Business School Working Paper 23-062; available at SSRN 4395751.

\bibitem[{Brown et~al.(2020)Brown, Mann, Ryder, Subbiah et~al.}]{BrownEtAl2020}
Brown TB, Mann B, Ryder N, Subbiah M, et~al. (2020) Language models are few-shot learners. \emph{Advances in Neural Information Processing Systems 33 (NeurIPS 2020)}, 1877--1901.

\bibitem[{Brynjolfsson et~al.(2011)Brynjolfsson, Hu, \protect\BIBand{} Simester}]{BrynjolfssonHuSimester2011}
Brynjolfsson E, Hu YJ, Simester D (2011) Goodbye {Pareto} principle, hello long tail: The effect of search costs on the concentration of product sales. \emph{Management Science} 57(8):1373--1386.

\bibitem[{Brynjolfsson et~al.(2003)Brynjolfsson, Hu, \protect\BIBand{} Smith}]{BrynjolfssonHuSmith2003}
Brynjolfsson E, Hu YJ, Smith MD (2003) Consumer surplus in the digital economy: Estimating the value of increased product variety at online booksellers. \emph{Management Science} 49(11):1580--1596.

\bibitem[{Brynjolfsson et~al.(2025)Brynjolfsson, Li, \protect\BIBand{} Raymond}]{BrynjolfssonLiRaymond2025}
Brynjolfsson E, Li D, Raymond LR (2025) Generative {AI} at work. \emph{Quarterly Journal of Economics} 140(2):889--942.

\bibitem[{Brynjolfsson \protect\BIBand{} Smith(2000)}]{BrynjolfssonSmith2000}
Brynjolfsson E, Smith MD (2000) Frictionless commerce? {A} comparison of {Internet} and conventional retailers. \emph{Management Science} 46(4):563--585.

\bibitem[{Buell \protect\BIBand{} Norton(2011)}]{BuellNorton2011}
Buell RW, Norton MI (2011) The labor illusion: How operational transparency increases perceived value. \emph{Management Science} 57(9):1564--1579.

\bibitem[{Cameron \protect\BIBand{} Huppert(1989)}]{CameronHuppert1989}
Cameron TA, Huppert DD (1989) {OLS} versus {ML} estimation of non-market resource values with payment card interval data. \emph{Journal of Environmental Economics and Management} 17(3):230--246.

\bibitem[{Cason \protect\BIBand{} Plott(2014)}]{CasonPlott2014}
Cason TN, Plott CR (2014) Misconceptions and game form recognition: Challenges to theories of revealed preference and framing. \emph{Journal of Political Economy} 122(6):1235--1270.

\bibitem[{Castelo et~al.(2019)Castelo, Bos, \protect\BIBand{} Lehmann}]{CasteloBosLehmann2019}
Castelo N, Bos MW, Lehmann DR (2019) Task-dependent algorithm aversion. \emph{Journal of Marketing Research} 56(5):809--825.

\bibitem[{Chen et~al.(2016)Chen, Schonger, \protect\BIBand{} Wickens}]{chen2016otree}
Chen DL, Schonger M, Wickens C (2016) otree—an open-source platform for laboratory, online, and field experiments. \emph{Journal of Behavioral and Experimental Finance} 9:88--97.

\bibitem[{Chen et~al.(2024)Chen, Kirshner, Ovchinnikov, Andiappan, \protect\BIBand{} Jenkin}]{ChenKirshnerEtAl2024}
Chen Y, Kirshner SN, Ovchinnikov A, Andiappan M, Jenkin T (2024) Biases of humans, of {AI}, and of humans with {AI} Working paper, available at SSRN 4943377.

\bibitem[{Chen et~al.(2025)Chen, Kirshner, Ovchinnikov, Andiappan, \protect\BIBand{} Jenkin}]{ChenKirshnerEtAl2025}
Chen Y, Kirshner SN, Ovchinnikov A, Andiappan M, Jenkin T (2025) A manager and an {AI} walk into a bar: Does {ChatGPT} make biased decisions like we do? \emph{Manufacturing \& Service Operations Management} 27(2):354--368.

\bibitem[{Chen \protect\BIBand{} Chan(2024)}]{ChenChan2024}
Chen Z, Chan J (2024) Large language model in creative work: The role of collaboration modality and user expertise. \emph{Management Science} 70(12):9101--9117.

\bibitem[{Chintagunta et~al.(2012)Chintagunta, Chu, \protect\BIBand{} Cebollada}]{ChintaguntaChuCebollada2012}
Chintagunta PK, Chu J, Cebollada J (2012) Quantifying transaction costs in online/off-line grocery channel choice. \emph{Marketing Science} 31(1):96--114.

\bibitem[{Choudhary(2026)}]{Choudhary2026RetailBrew}
Choudhary V (2026) Why agentic shopping adoption remains low despite the hype. Retail Brew.

\bibitem[{Cinelli et~al.(2021)Cinelli, De~Francisci~Morales, Galeazzi, Quattrociocchi, \protect\BIBand{} Starnini}]{cinelli2021echo}
Cinelli M, De~Francisci~Morales G, Galeazzi A, Quattrociocchi W, Starnini M (2021) The echo chamber effect on social media. \emph{Proceedings of the National Academy of Sciences} 118(9):e2023301118, \urlprefix\url{http://dx.doi.org/10.1073/pnas.2023301118}.

\bibitem[{Dell'Acqua et~al.(2023)Dell'Acqua, McFowland~III, Mollick, Lifshitz-Assaf, Kellogg, Rajendran, Krayer, Candelon, \protect\BIBand{} Lakhani}]{DellAcquaEtAl2023}
Dell'Acqua F, McFowland~III E, Mollick ER, Lifshitz-Assaf H, Kellogg K, Rajendran S, Krayer L, Candelon F, Lakhani KR (2023) Navigating the jagged technological frontier: Field experimental evidence of the effects of {AI} on knowledge worker productivity and quality. Technical Report Working Paper 24-013, Harvard Business School.

\bibitem[{Dhar \protect\BIBand{} Wertenbroch(2000)}]{DharWertenbroch2000}
Dhar R, Wertenbroch K (2000) Consumer choice between hedonic and utilitarian goods. \emph{Journal of Marketing Research} 37(1):60--71.

\bibitem[{Dietvorst et~al.(2015)Dietvorst, Simmons, \protect\BIBand{} Massey}]{DietvorstSimmonsMassey2015}
Dietvorst BJ, Simmons JP, Massey C (2015) Algorithm aversion: People erroneously avoid algorithms after seeing them err. \emph{Journal of Experimental Psychology: General} 144(1):114--126.

\bibitem[{Ding et~al.(2005)Ding, Grewal, \protect\BIBand{} Liechty}]{DingGrewalLiechty2005}
Ding M, Grewal R, Liechty J (2005) Incentive-aligned conjoint analysis. \emph{Journal of Marketing Research} 42(1):67--82.

\bibitem[{DosSantos~DiSorbo et~al.(2025)DosSantos~DiSorbo, Ferreira, Balakrishnan, \protect\BIBand{} Tong}]{DiSorboFerreiraBalakrishnanTong2025}
DosSantos~DiSorbo M, Ferreira KJ, Balakrishnan M, Tong J (2025) Warnings and endorsements: Improving human--ai collaboration in the presence of outliers. \emph{Manufacturing \& Service Operations Management} 27(6):1814--1831.

\bibitem[{Gao et~al.(2023)Gao, Xiong, Gao, Jia, Pan, Bi, Dai, Sun, Wang, \protect\BIBand{} Wang}]{GaoEtAl2023}
Gao Y, Xiong Y, Gao X, Jia K, Pan J, Bi Y, Dai Y, Sun J, Wang H, Wang M (2023) Retrieval-augmented generation for large language models: A survey. \emph{arXiv preprint arXiv:2312.10997} .

\bibitem[{Girotra et~al.(2023)Girotra, Meincke, Terwiesch, \protect\BIBand{} Ulrich}]{GirotraEtAl2023}
Girotra K, Meincke L, Terwiesch C, Ulrich KT (2023) Ideas are dimes a dozen: Large language models for idea generation in innovation Working paper, The Wharton School, University of Pennsylvania.

\bibitem[{Hagendorff et~al.(2023)Hagendorff, Fabi, \protect\BIBand{} Kosinski}]{HagendorffEtAl2023}
Hagendorff T, Fabi S, Kosinski M (2023) Human-like intuitive behavior and reasoning biases emerged in large language models but disappeared in {ChatGPT}. \emph{Nature Computational Science} 3:833--838.

\bibitem[{Hann \protect\BIBand{} Terwiesch(2003)}]{HannTerwiesch2003}
Hann IH, Terwiesch C (2003) Measuring the frictional costs of online transactions: The case of a name-your-own-price channel. \emph{Management Science} 49(11):1563--1579.

\bibitem[{Harrison \protect\BIBand{} Rutstr{\"o}m(2008)}]{HarrisonRutstrom2008}
Harrison GW, Rutstr{\"o}m EE (2008) Risk aversion in the laboratory. Cox JC, Harrison GW, eds., \emph{Risk Aversion in Experiments}, volume~12 of \emph{Research in Experimental Economics}, 41--196 (Emerald Group Publishing).

\bibitem[{H\"aubl \protect\BIBand{} Trifts(2000)}]{HaublTrifts2000}
H\"aubl G, Trifts V (2000) Consumer decision making in online shopping environments: The effects of interactive decision aids. \emph{Marketing Science} 19(1):4--21.

\bibitem[{Hauser \protect\BIBand{} Wernerfelt(1990)}]{HauserWernerfelt1990}
Hauser JR, Wernerfelt B (1990) An evaluation cost model of consideration sets. \emph{Journal of Consumer Research} 16(4):393--408.

\bibitem[{Holt \protect\BIBand{} Laury(2002)}]{HoltLaury2002}
Holt CA, Laury SK (2002) Risk aversion and incentive effects. \emph{American Economic Review} 92(5):1644--1655.

\bibitem[{Homburg et~al.(2005)Homburg, Koschate, \protect\BIBand{} Hoyer}]{homburg2005satisfied}
Homburg C, Koschate N, Hoyer WD (2005) Do satisfied customers really pay more? a study of the relationship between customer satisfaction and willingness to pay. \emph{Journal of marketing} 69(2):84--96.

\bibitem[{Horton(2023)}]{Horton2023}
Horton JJ (2023) Large language models as simulated economic agents: What can we learn from {Homo Silicus}? NBER Working Paper 31122.

\bibitem[{Johnson et~al.(2021)Johnson, Douze, \protect\BIBand{} J\'egou}]{johnson2019billion}
Johnson J, Douze M, J\'egou H (2021) Billion-scale similarity search with {GPUs}. \emph{IEEE Transactions on Big Data} 7(3):535--547.

\bibitem[{Kagan et~al.(2026)Kagan, Hathaway, \protect\BIBand{} Dada}]{KaganHathawayDada2026}
Kagan E, Hathaway B, Dada M (2026) Why are customers averse to service chatbots? \emph{Manufacturing \& Service Operations Management} Forthcoming.

\bibitem[{Kremer et~al.(2011)Kremer, Moritz, \protect\BIBand{} Siemsen}]{KremerMoritzSiemsen2011}
Kremer M, Moritz B, Siemsen E (2011) Demand forecasting behavior: System neglect and change detection. \emph{Management Science} 57(10):1827--1843.

\bibitem[{Lee \protect\BIBand{} Donohue(2026)}]{LeeDonohue2026}
Lee Y, Donohue K (2026) Nudging green-but-slow shipping choices in online retail. \emph{Management Science} Articles in Advance.

\bibitem[{Lewis et~al.(2020)Lewis, Perez, Piktus, Petroni, Karpukhin, Goyal, K\"uttler, Lewis, Yih, Rockt\"aschel, Riedel, \protect\BIBand{} Kiela}]{LewisEtAl2020RAG}
Lewis P, Perez E, Piktus A, Petroni F, Karpukhin V, Goyal N, K\"uttler H, Lewis M, Yih Wt, Rockt\"aschel T, Riedel S, Kiela D (2020) Retrieval-augmented generation for knowledge-intensive {NLP} tasks. \emph{Advances in Neural Information Processing Systems (NeurIPS)}, volume~33, 9459--9474.

\bibitem[{Lin et~al.(2025)Lin, Shi, \protect\BIBand{} Sun}]{LinShiSun2025}
Lin S, Shi Z, Sun X (2025) Towards intelligent shopping assistant: How can {LLM}-based chatbots transform online shopping process? Working paper; available at SSRN 5088975.

\bibitem[{Logg et~al.(2019)Logg, Minson, \protect\BIBand{} Moore}]{LoggMinsonMoore2019}
Logg JM, Minson JA, Moore DA (2019) Algorithm appreciation: People prefer algorithmic to human judgment. \emph{Organizational Behavior and Human Decision Processes} 151:90--103.

\bibitem[{Longoni et~al.(2019)Longoni, Bonezzi, \protect\BIBand{} Morewedge}]{LongoniBonezziMorewedge2019}
Longoni C, Bonezzi A, Morewedge CK (2019) Resistance to medical artificial intelligence. \emph{Journal of Consumer Research} 46(4):629--650.

\bibitem[{Longoni \protect\BIBand{} Cian(2022)}]{LongoniCian2022}
Longoni C, Cian L (2022) Artificial intelligence in utilitarian vs. hedonic contexts: The ``word-of-machine'' effect. \emph{Journal of Marketing} 86(1):91--108.

\bibitem[{Luo et~al.(2019)Luo, Tong, Fang, \protect\BIBand{} Qu}]{LuoTongFangQu2019}
Luo X, Tong S, Fang Z, Qu Z (2019) Frontiers: Machines vs. humans: The impact of artificial intelligence chatbot disclosure on customer purchases. \emph{Marketing Science} 38(6):937--947.

\bibitem[{Mackworth(1948)}]{mackworth1948breakdown}
Mackworth NH (1948) The breakdown of vigilance during prolonged visual search. \emph{Quarterly journal of experimental psychology} 1(1):6--21.

\bibitem[{Maister(1984)}]{maister1984}
Maister D (1984) \emph{The psychology of waiting lines} (Harvard Business School Boston).

\bibitem[{Marchionini(2006)}]{Marchionini2006}
Marchionini G (2006) Exploratory search: From finding to understanding. \emph{Communications of the ACM} 49(4):41--46.

\bibitem[{Moe(2003)}]{Moe2003}
Moe WW (2003) {Buying, Searching, or Browsing: Differentiating Between Online Shoppers Using In-Store Navigational Clickstream}. \emph{Journal of Consumer Psychology} 13(1--2):29--39.

\bibitem[{Moritz et~al.(2014)Moritz, Siemsen, \protect\BIBand{} Kremer}]{MoritzSiemsenKremer2014}
Moritz B, Siemsen E, Kremer M (2014) Judgmental forecasting: Cognitive reflection and decision speed. \emph{Production and Operations Management} 23(7):1146--1160.

\bibitem[{Ni et~al.(2026)Ni, Wang, Feng, Lu, Wang, \protect\BIBand{} Zhou}]{NiEtAl2026}
Ni X, Wang Y, Feng T, Lu LX, Wang Y, Zhou C (2026) Generative {AI} in action: Field experimental evidence from {Alibaba}'s customer service operations ArXiv preprint arXiv:2603.29888; available at SSRN 5012601.

\bibitem[{Noy \protect\BIBand{} Zhang(2023)}]{NoyZhang2023}
Noy S, Zhang W (2023) Experimental evidence on the productivity effects of generative artificial intelligence. \emph{Science} 381(6654):187--192.

\bibitem[{{NVIDIA}(2023)}]{NVIDIA2023RAG}
{NVIDIA} (2023) What is retrieval-augmented generation, aka {RAG}? NVIDIA Blog.

\bibitem[{Peng et~al.(2023)Peng, Kalliamvakou, Cihon, \protect\BIBand{} Demirer}]{PengEtAl2023}
Peng S, Kalliamvakou E, Cihon P, Demirer M (2023) The impact of {AI} on developer productivity: Evidence from {GitHub} {Copilot} ArXiv preprint arXiv:2302.06590.

\bibitem[{Puntoni et~al.(2021)Puntoni, Reczek, Giesler, \protect\BIBand{} Botti}]{PuntoniEtAl2021}
Puntoni S, Reczek RW, Giesler M, Botti S (2021) Consumers and artificial intelligence: An experiential perspective. \emph{Journal of Marketing} 85(1):131--151.

\bibitem[{Radlinski \protect\BIBand{} Craswell(2017)}]{RadlinskiCraswell2017}
Radlinski F, Craswell N (2017) A theoretical framework for conversational search. \emph{Proceedings of the 2017 Conference on Human Information Interaction and Retrieval}, 117--126 (Association for Computing Machinery).

\bibitem[{Reisenbichler et~al.(2022)Reisenbichler, Reutterer, Schweidel, \protect\BIBand{} Dan}]{ReisenbichlerEtAl2022}
Reisenbichler M, Reutterer T, Schweidel DA, Dan D (2022) Frontiers: Supporting content marketing with natural language generation. \emph{Marketing Science} 41(3):441--452.

\bibitem[{Schweitzer \protect\BIBand{} Cachon(2000)}]{SchweitzerCachon2000}
Schweitzer ME, Cachon GP (2000) Decision bias in the newsvendor problem with a known demand distribution: Experimental evidence. \emph{Management Science} 46(3):404--420.

\bibitem[{Sharma et~al.(2024)Sharma, Liao, \protect\BIBand{} Xiao}]{sharma2024}
Sharma N, Liao QV, Xiao Z (2024) Generative echo chamber? effect of llm-powered search systems on diverse information seeking. \emph{Proceedings of the 2024 CHI Conference on Human Factors in Computing Systems}, 1--17.

\bibitem[{Shuster et~al.(2021)Shuster, Poff, Chen, Kiela, \protect\BIBand{} Weston}]{ShusterEtAl2021}
Shuster K, Poff S, Chen M, Kiela D, Weston J (2021) Retrieval augmentation reduces hallucination in conversation. \emph{Findings of the Association for Computational Linguistics: EMNLP 2021}, 3784--3803.

\bibitem[{Slovic(1995)}]{Slovic1995}
Slovic P (1995) The construction of preference. \emph{American Psychologist} 50(5):364--371.

\bibitem[{Song et~al.(2025)Song, Gaur, \protect\BIBand{} Zhu}]{SongGaurZhu2025}
Song B, Gaur V, Zhu R (2025) Can {LLM} chatbots replace flow-driven chatbots in customer service? Cornell SC Johnson College of Business Research Paper, forthcoming; available at SSRN.

\bibitem[{Spatharioti et~al.(2025)Spatharioti, Rothschild, Goldstein, \protect\BIBand{} Hofman}]{SpathariotiEtAl2025}
Spatharioti SE, Rothschild DM, Goldstein DG, Hofman JM (2025) Effects of {LLM}-based search on decision making: Speed, accuracy, and overreliance. \emph{Proceedings of the 2025 CHI Conference on Human Factors in Computing Systems (CHI '25)} (New York, NY, USA: ACM).

\bibitem[{Stigler(1961)}]{Stigler1961}
Stigler GJ (1961) The economics of information. \emph{Journal of Political Economy} 69(3):213--225.

\bibitem[{Subramonyam et~al.(2024)Subramonyam, Im, Seifert, \protect\BIBand{} Adar}]{SubramonyamEtAl2024}
Subramonyam H, Im J, Seifert C, Adar E (2024) Bridging the gulf of envisioning: Cognitive challenges in prompt based interactions with {LLMs}. \emph{Proceedings of the 2024 CHI Conference on Human Factors in Computing Systems (CHI '24)} (New York, NY, USA: ACM).

\bibitem[{Suri et~al.(2024)Suri, Slater, Ziaee, \protect\BIBand{} Nguyen}]{SuriEtAl2024}
Suri G, Slater LR, Ziaee A, Nguyen M (2024) Do large language models show decision heuristics similar to humans? a case study using {GPT-3.5}. \emph{Journal of Experimental Psychology: General} 153(4):1066--1075.

\bibitem[{Wang et~al.(2025)Wang, Zhang, \protect\BIBand{} Zhang}]{WangZhangZhang2025}
Wang M, Zhang DJ, Zhang H (2025) Large language models for market research: A data-augmentation approach. \emph{Marketing Science} Forthcoming.

\bibitem[{Wei et~al.(2022)Wei, Bosma, Zhao, Guu, Yu, Lester, Du, Dai, \protect\BIBand{} Le}]{WeiEtAl2022}
Wei J, Bosma M, Zhao VY, Guu K, Yu AW, Lester B, Du N, Dai AM, Le QV (2022) Finetuned language models are zero-shot learners. \emph{International Conference on Learning Representations (ICLR)}.

\bibitem[{Weitzman(1979)}]{Weitzman1979}
Weitzman ML (1979) Optimal search for the best alternative. \emph{Econometrica} 47(3):641--654.

\bibitem[{Wertenbroch \protect\BIBand{} Skiera(2002)}]{WertenbrochSkiera2002}
Wertenbroch K, Skiera B (2002) {Measuring Consumers' Willingness to Pay at the Point of Purchase}. \emph{Journal of Marketing Research} 39(2):228--241.

\bibitem[{Zamfirescu-Pereira et~al.(2023)Zamfirescu-Pereira, Wong, Hartmann, \protect\BIBand{} Yang}]{ZamfirescuPereiraEtAl2023}
Zamfirescu-Pereira JD, Wong RY, Hartmann B, Yang Q (2023) Why {J}ohnny can't prompt: How non-{AI} experts try (and fail) to design {LLM} prompts. \emph{Proceedings of the 2023 CHI Conference on Human Factors in Computing Systems (CHI '23)} (New York, NY, USA: ACM).

\bibitem[{Zhang \protect\BIBand{} Narayandas(2025)}]{ZhangNarayandas2025}
Zhang S, Narayandas D (2025) Engaging customers with {AI} in online chats: Evidence from a randomized field experiment. \emph{Management Science} Forthcoming.

\end{thebibliography}
 
\newpage
\clearpage

\ECHead{Electronic Companion}
\small

\renewcommand{\thesection}{EC.\arabic{section}}
\renewcommand{\thesubsection}{\thesection.\arabic{subsection}}
\renewcommand{\thefigure}{\thesection.\arabic{figure}}
\renewcommand{\thetable}{\thesection.\arabic{table}}
\renewcommand{\theHsection}{EC.\arabic{section}}
\renewcommand{\theHsubsection}{\theHsection.\arabic{subsection}}
\renewcommand{\theHsubsubsection}{\theHsubsection.\arabic{subsubsection}}
\renewcommand{\theHfigure}{EC.\arabic{section}.\arabic{figure}}
\renewcommand{\theHtable}{EC.\arabic{section}.\arabic{table}}
\setcounter{section}{0}
\setcounter{subsection}{0}
\setcounter{figure}{0}
\setcounter{table}{0}
\makeatletter
\@addtoreset{figure}{section}
\@addtoreset{table}{section}
\@addtoreset{subsection}{section}
\makeatother

\noindent This companion includes the following:
\smallskip

\begin{itemize}[leftmargin=1.8cm, labelwidth=0.9cm, labelsep=0.3cm, align=left, itemsep=4pt]
    \item[EC.1] \hyperref[app:rag]{RAG architecture, system prompts, and retrieval-accuracy validation} (Supplementing \textsection\ref{subsec:llm-customization} and \textsection\ref{sec:meth-rag})
    \item[EC.2] \hyperref[app:balance]{Balance tests} (Supplementing \textsection\ref{sec:meth-overview})
    \item[EC.3] \hyperref[app:pilots]{Pilot studies} (Supplementing \textsection\ref{sec:meth-catalog})
    \item[EC.4] \hyperref[app:materials]{Experimental interface and instructions} (Supplementing \textsection\ref{sec:meth-procedure})
    \item[EC.5] \hyperref[app:model-intent]{Model performance and search intent} (Supplementing \textsection\ref{sec:res-model-intent-model})
    \item[EC.6] \hyperref[app:mech]{Mechanism Analysis} (Supplementing \textsection\ref{sec:Mechanisms})

\end{itemize}

\newpage

\section{Retrieval-Augmented Generation (RAG): Architecture, System Prompts, and Validation}\label{app:rag}

This appendix supplements \textsection\ref{sec:meth-rag} with details on the RAG implementation. \textsection\ref{app:rag-choice} discusses our choice of RAG relative to the alternative approaches for customizing an LLM. \textsection\ref{app:rag-pipeline} describes the pipeline in detail. \textsection\ref{app:rag-schematic} provides diagrams for each of the three RAG stages. \textsection\ref{app:rag-embedding} illustrates retrieval in the embedding space with two example queries. \textsection\ref{app:rag-prompts} reproduces the verbatim system prompts of both LLMs. \textsection\ref{app:rag-mc} reports a Monte Carlo simulation that validates the system's retrieval accuracy.

\subsection{Choice of Customization Approach}\label{app:rag-choice}

RAG is one of three standard approaches for customizing an LLM; the others are \emph{system prompting} and \emph{fine-tuning}. The main difference between these approaches is in how they treat the model's \emph{weights} (the numerical parameters set during pretraining that determine the model's outputs). Table~\ref{tab:deployment-approaches} compares the three approaches on the properties relevant to our study: the amount of factual knowledge they add, whether they can be updated without retraining, and whether they expose a retrieval step that can be inspected. System prompting puts instructions and context inside the prompt \citep{BrownEtAl2020}. While this approach is flexible and allows us (in principle) to achieve our goals of adding event information to the LLM's knowledge base, the context window is typically limited in terms of how much information the prompt can hold. Fine-tuning updates the weights on labeled examples \citep{WeiEtAl2022}, which means that the knowledge it adds is stored in the weights and can be changed only by retraining. Furthermore, the retrieval step is not verifiable, as the new information is essentially blended with the original training data of the LLM in a ``black-box'' manner. RAG pairs the LLM with an external \emph{knowledge base} and retrieves matching entries at query time \citep{LewisEtAl2020RAG, GaoEtAl2023}. RAG can add a large amount of factual knowledge, is updated by re-indexing the knowledge base, and produces a retrieval step that can be inspected and validated separately from the LLM, which we do in our analysis.

We choose RAG because our application requires: (i) accurate knowledge of a specific catalog of events, (ii) the ability to refresh the catalog across data-collection waves without retraining, and (iii) a retrieval step that we can inspect and validate independently of the LLM (\textsection\ref{app:rag-mc}). Prompting does not satisfy (i) at our scale: placing thousands of events in every prompt is costly, slows each reply, and degrades retrieval quality, and a refreshed catalog can exceed the model's context window. Fine-tuning satisfies neither (ii) nor (iii), since updating the catalog would require retraining and the model exposes no separable retrieval step to audit. Notably, system prompting still plays a role in our design, but only for instructions and tone (\textsection\ref{app:rag-prompts}); it is not how the LLM learns the catalog. 

\begin{table}[hb]
\TABLE
{Comparison of LLM Customization Approaches\label{tab:deployment-approaches}}
{\footnotesize
\begin{tabular}{lccc}
\hline
                                       & System Prompting           & Fine-tuning           & RAG                 \\
\hline
Changes the model's weights?           & No                  & Yes                   & No                  \\
Adds new factual knowledge?            & Limited             & Not reliably          & Yes                 \\
Best suited for\dots                   & Instructions, tone  & Consistent behavior   & Domain knowledge    \\
Updatable without retraining?         & Yes                 & No                    & Yes (re-index)      \\
Inspectable retrieval step?            & ---                 & No                    & Yes                 \\
\hline
\end{tabular}}
{}
\end{table}

\subsection{Pipeline}\label{app:rag-pipeline}

The RAG service runs as a separate FastAPI process. The oTree front end posts each user message to the service via a JSON endpoint and receives back a natural-language reply together with a list of event ids to render in the panel below the chat. The service exposes a single \texttt{/query} endpoint that takes the user message, the recent conversation history, and an optional model identifier (used to randomize the generation model across participants in the \textit{Chat} arm), and returns the model's reply together with the event ids it selected.

Internally, the service runs the following four stages.

\begin{enumerate}[label=(\roman*),leftmargin=*]
\item \textbf{Indexing.} On startup, the service loads the experimental catalog (a CSV with one event per row) and computes an embedding of each event's textual record. Each record is a key-value serialization of the event's structured attributes (event id, name, date, time, venue, NYC borough, segment, genre, subgenre).

A standard RAG deployment requires the researcher to choose how to split the source document(s) into smaller indexable units (``chunks''); typical settings include chunk size (200--500 tokens), chunk overlap (10--20\% of chunk size), and the number of chunks to retrieve per query \citep{GaoEtAl2023}. In our setting, this design problem collapses. The source ``document'' is a single CSV file with one row per event, and each row is short ($<200$ tokens) and self-contained. We therefore treat each row as a single indexable record; chunk size, overlap, and chunk boundaries are all dictated by the structure of the data rather than chosen by the researcher. The only retrieval parameters that remain to be tuned are the similarity threshold and top-$k$, both of which we calibrated in the pilots (\textsection\ref{app:pilots}).

We use OpenAI's \texttt{text-embedding-ada-002} model to embed each record. The embedding vectors are unit-normalized, so nearest neighbors by $L_2$ distance coincide with nearest neighbors by cosine similarity. The resulting vectors are stored in a FAISS \citep{johnson2019billion} index that is persisted to disk and reused across sessions; the catalog is hashed and the index is rebuilt only when the catalog changes. The index is identical across all participants and across all model conditions.

\item \textbf{Routing.} When the front end posts a user message, the service first invokes a deterministic \emph{router} LLM (\texttt{gpt-5.2}, temperature $0.1$, max $800$ tokens). The router LLM has access to the same retrieval tool described below as a function call but is instructed to invoke it only when the user message is an event-search request. For greetings or thanks, the router replies conversationally; for vague requests, it asks a clarifying question; for unrelated questions, it returns a fixed redirection. The router is held fixed across all model conditions, so the binary decision of whether to retrieve, and (if so) the rewritten query passed to retrieval, are identical regardless of which generation model is used downstream.

\item \textbf{Retrieval.} If the router invokes the retrieval tool, the service embeds the rewritten query with the same embedding model used at indexing time and queries the FAISS index. We retain all candidates whose $L_2$ distance to the query embedding is at most $1.2$ (chosen empirically during pilots to balance recall and precision). The retained set is then narrowed by a deterministic temporal pre-filter: any day-of-week or month names mentioned in the query (parsed via a regular-expression dictionary) are converted to date constraints, and events whose dates fall outside those constraints are dropped. The result is then capped at the top $500$ events ranked by similarity. If the LLM passes an explicit date in YYYY-MM-DD format as a tool argument, the similarity cutoff is bypassed and the entire catalog is filtered exactly to that date instead.

\item \textbf{Generation.} The retained candidates, the user message, and the last six conversational turns are passed to the \emph{generation} LLM (the model that varies across participants in the \textit{Chat} arm: \texttt{gpt-4.1-mini}, \texttt{gpt-5.2}, or \texttt{gpt-5.4}; temperature $0.2$, max $800$ tokens, $30$-second timeout). The generation system prompt instructs the model to ground its reply in the supplied events, to keep the reply concise (one to two sentences), to never mention prices, and, importantly, to generate a tagged list of event ids it considers relevant at the end of its response. The service then parses this tagged list, intersects it with the catalog's id set (so any id the model invents is silently dropped), caps the list at $100$ ids, and returns the cleaned reply and id list to the front end. The front end renders the corresponding events in a paginated list of ten per page.
\end{enumerate}

\subsection{Pipeline Diagrams}\label{app:rag-schematic}

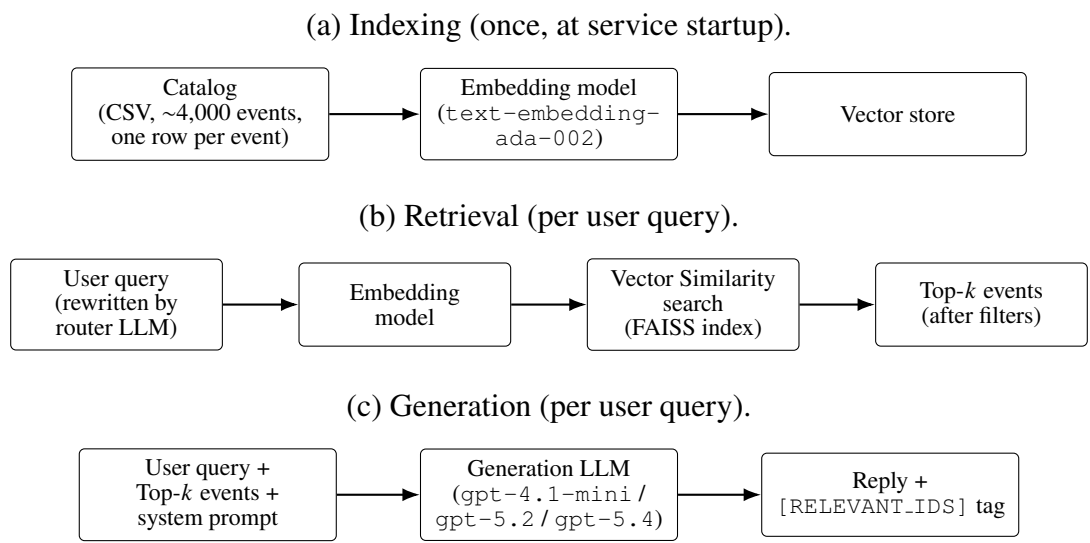
\begin{figure}[h]
\caption{RAG architecture: the three stages of the pipeline.}
\label{fig:rag-schematic}
\centering
\begin{subfigure}{\linewidth}
\centering
\caption{Indexing (once, at service startup).}
\label{fig:rag-indexing}
\begin{tikzpicture}[
  node distance=1.2cm,
  box/.style={draw, rectangle, rounded corners=2pt, minimum width=3.4cm, minimum height=1.1cm, align=center, font=\footnotesize\linespread{0.85}\selectfont},
  arr/.style={-Latex, thick}
]
\node[box] (cat) {Catalog\\(CSV, $\sim$$4{,}000$ events,\\one row per event)};
\node[box, right=of cat] (emb) {Embedding model\\(\texttt{text-embedding-}\\\texttt{ada-002})};
\node[box, right=of emb] (vs)  {Vector store};
\draw[arr] (cat) -- (emb);
\draw[arr] (emb) -- (vs);
\end{tikzpicture}
\end{subfigure}\\[10pt]

\begin{subfigure}{\linewidth}
\centering
\caption{Retrieval (per user query).}
\label{fig:rag-retrieval}
\begin{tikzpicture}[
  node distance=1.0cm,
  box/.style={draw, rectangle, rounded corners=2pt, minimum width=2.8cm, minimum height=1.1cm, align=center, font=\footnotesize\linespread{0.85}\selectfont},
  arr/.style={-Latex, thick}
]
\node[box] (q) {User query\\(rewritten by\\router LLM)};
\node[box, right=of q] (qemb) {Embedding\\model};
\node[box, right=of qemb] (search) {Vector Similarity\\search \\(FAISS index)};
\node[box, right=of search] (out) {Top-$k$ events\\(after filters)};
\draw[arr] (q) -- (qemb);
\draw[arr] (qemb) -- (search);
\draw[arr] (search) -- (out);
\end{tikzpicture}
\end{subfigure}\\[10pt]

\begin{subfigure}{\linewidth}
\centering
\caption{Generation (per user query).}
\label{fig:rag-generation}
\begin{tikzpicture}[
  node distance=1.1cm,
  box/.style={draw, rectangle, rounded corners=2pt, minimum width=3.4cm, minimum height=1.1cm, align=center, font=\footnotesize\linespread{0.85}\selectfont},
  arr/.style={-Latex, thick}
]
\node[box] (in) {User query +\\Top-$k$ events +\\system prompt};
\node[box, right=of in] (llm) {Generation LLM\\(\texttt{gpt-4.1-mini} /\\ \texttt{gpt-5.2} / \texttt{gpt-5.4})};
\node[box, right=of llm] (out) {Reply +\\\texttt{[RELEVANT\_IDS]} tag};
\draw[arr] (in)  -- (llm);
\draw[arr] (llm) -- (out);
\end{tikzpicture}
\end{subfigure}

\vspace{3pt}

\begin{minipage}{0.89\linewidth}
\vspace{3pt}

\footnotesize\linespread{0.9}\selectfont \emph{Note.} Indexing runs once at service startup; retrieval and generation run on every user message. \textsection\ref{app:rag-pipeline} details each stage, including the router LLM that precedes retrieval. \par
\end{minipage}
\end{figure}

Figure~\ref{fig:rag-schematic} summarizes the three RAG stages. 

\subsection{Retrieval in the Embedding Space: Two Examples}\label{app:rag-embedding}

\begin{figure}
\FIGURE
{\includegraphics[width=\linewidth]{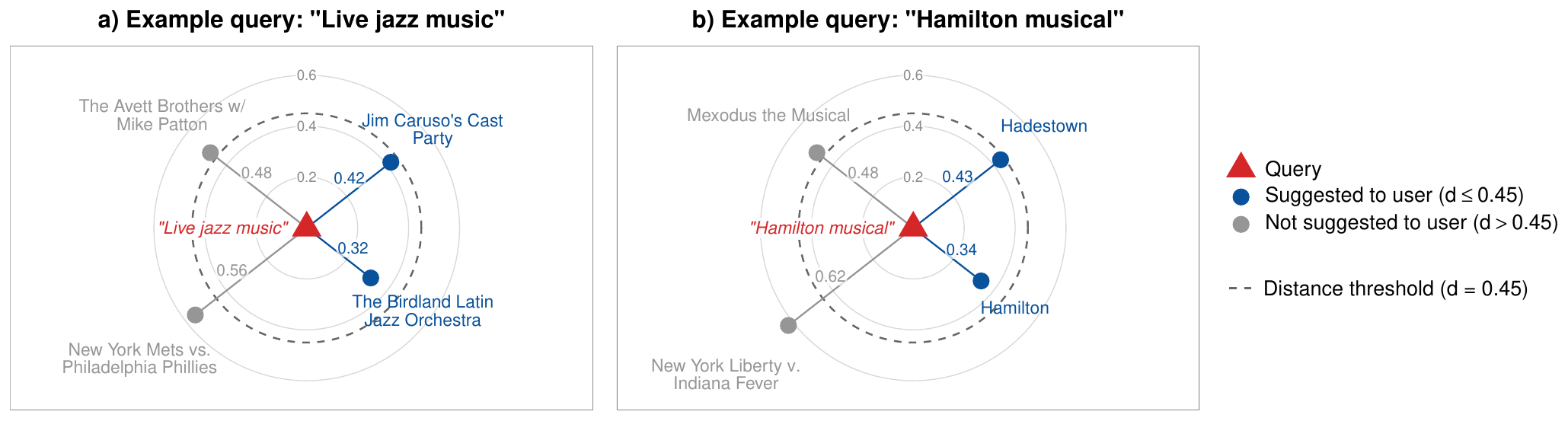}}
{Retrieval as nearest-neighbor search in the embedding space\label{fig:embed-space}}
{}
\end{figure}

Figure~\ref{fig:embed-space} illustrates how retrieval works, using two example queries in panels (a) and (b). The triangle at the center represents the user query and the dashed circle in each panel marks the retrieval cutoff of the $L_{2}$ distance, i.e., events within the dashed circle are retrieved and shown to the user. For this illustration, we set this distance to $0.45$.\footnote{In the experiment we use a more generous cutoff of 1.2. Because results are paginated (ten results per page), weaker matches can still be included in the results rather than being dropped, so a user who keeps browsing and scrolling through the pages can still reach them.} In panel (a), a user searches for \emph{``live jazz music''}. A traditional search box would return only events whose names contain those words: for example, it would find the event \emph{The Birdland Latin Jazz Orchestra}. Our RAG-based AI search also finds this event and assigns it a relatively small distance. However, because AI search matches based on meaning (rather than on string matches), it also returns events like \emph{Jim Caruso's Cast Party}, a weekly live-music night at a New York jazz club that shares no words with the query but that the embedding has learned to associate with it. Showing these types of matches is the main advantage of AI-based search over a string-based filter.

Panel (b) of Figure~\ref{fig:embed-space} shows that the same mechanism can exclude an event that would be included with a string-based search but that does not match well the intent of the query. In particular, for \emph{``Hamilton musical''}, the system retrieves \emph{Hadestown}, which shares no words with the query, but omits \emph{Mexodus the Musical}, even though \emph{Mexodus} matches both the segment (Arts \& Theater) and the word ``musical.'' This is because \emph{Hadestown}, like \emph{Hamilton}, is a prominent Broadway show, whereas \emph{Mexodus} is an off-Broadway show and therefore has a higher $L_{2}$ distance.

\subsection{System Prompts}\label{app:rag-prompts}

The prompts reproduced below are those used in Wave 2, which served events scheduled in May and June 2026. The Wave 1 prompts were identical except for the month references (``April and May'' rather than ``May and June'').

The router LLM is instantiated with the following system prompt:

\begin{verbatimbox}\footnotesize\ttfamily
You are an event-search assistant for NYC events in May and June 2026. If the user is asking about events, entertainment, or things to do, call the retrieve tool. If the user sends a greeting or thanks (e.g.\ `hi', `hello', `thanks'), respond conversationally WITHOUT calling retrieve. If the query is vague (e.g.\ `fun things', `something cool', `what should I do'), ask a brief clarifying question WITHOUT calling retrieve.

\smallskip
REFINEMENTS \& FOLLOW-UPS: When the user refines a previous search (e.g.\ `sports' after asking about a date, `not jazz', `only comedy', `remove the classical ones'), ALWAYS call retrieve again with a combined query that captures the FULL intent. For example: previous query `may 6' + follow-up `sports' $\to$ retrieve(`sports events on may 6'). Previous query `jazz' + follow-up `not blues' $\to$ retrieve(`jazz events excluding blues'). Always combine the conversation context into a single clear retrieve query.

\smallskip
For ANY other non-event question, respond ONLY with: ``I can only help you find events. What events are you interested in?''. Do NOT answer general knowledge questions, trivia, opinions, or anything unrelated to events.
\end{verbatimbox}

The generation LLM is instantiated with the following system prompt:

\begin{verbatimbox}\footnotesize\ttfamily
You are an event search assistant for NYC events in May and June 2026. Use the provided events to answer the user's question. NEVER invent events or details not in the data. NEVER mention anything related to prices, money, budget, cost, or fees. Be concise (1--2 sentences max). The user can see the events in a panel below---do NOT list specific events in your response. Do NOT list event names, dates, or venues. Just summarize what you found. Refer the user to the events panel to browse details.

\smallskip
IMPORTANT: From the candidate events, decide which ones are RELEVANT to the user's current request. At the very end of your response, on a new line, output exactly: \texttt{[RELEVANT\_IDS: id1, id2, id3, \dots]} listing only the event\_ids that match. Include at most 100 IDs. If the user asks for exclusions (e.g.\ `not latin', `only knicks'), apply those strictly. When the user specifies a date range, ONLY include events within that exact range. Do NOT include events outside the requested dates. This tag controls which events the user sees---missing a relevant event means the user won't see it.
\end{verbatimbox}

\subsection{Validation: Monte Carlo Retrieval Accuracy}\label{app:rag-mc}

In this section, we validate the retrieval quality of the system via a Monte Carlo simulation against each wave's catalog. In each of $1{,}000$ iterations, we sample one event uniformly at random from the catalog and construct a structured natural-language query of the form ``$\langle$segment$\rangle$ event on $\langle$date$\rangle$ $\langle$keyword$\rangle$,'' where the date is the event's calendar date and the keyword is one or two distinctive words extracted from the event title (subtitles after a colon or em-dash are stripped, and stopwords are excluded). We then submit the same query to all three retrieval systems and record whether the sampled event's id appears (i) anywhere in the returned result set (``in set'') and (ii) within the first ten results (``top-10''). The three systems mirror their main-experiment counterparts: \textit{Filter} applies a deterministic substring filter on segment, date, and keyword over the catalog; \textit{Chat} posts the full query to the live RAG service and uses the model's tagged \texttt{[RELEVANT\_IDS]} list; and \textit{Hybrid} returns the union of the two id streams.

\begin{figure}[h]
\caption{Monte Carlo retrieval accuracy by treatment.}
\label{fig:mc-accuracy}
\centering
\includegraphics[width=0.95\linewidth]{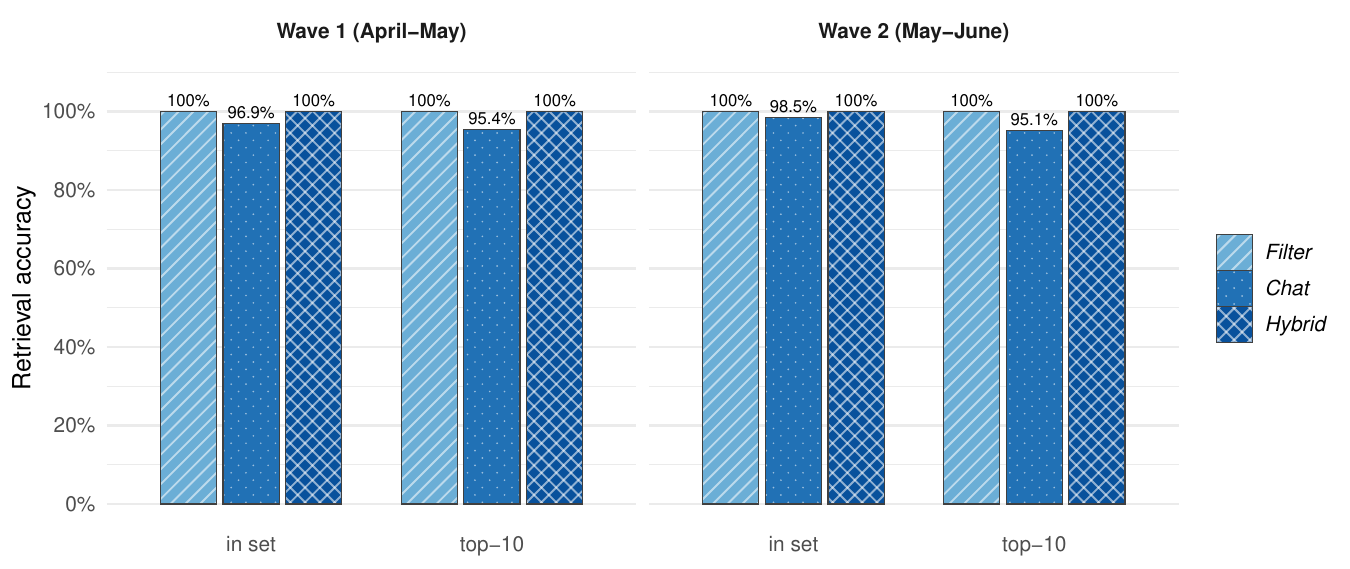}
\flushleft
\footnotesize \emph{Note.} Each iteration draws one event at random from the wave's catalog and submits a structured natural-language query (segment, date, and one or two distinctive title words) to each of the three systems. ``In set'' records whether the target event appears anywhere in the system's result list; ``top-10'' restricts to the first ten results. \textit{Filter} is the deterministic substring filter; \textit{Chat} is the live RAG service (model: \texttt{gpt-5.2}); \textit{Hybrid} returns the union of the two. Both waves use $N = 1{,}000$ iterations.
\end{figure}

Figure~\ref{fig:mc-accuracy} summarizes the results. The deterministic filter recovers every sampled event in both metrics across both catalogs, as does the hybrid system. The conversational system retrieves the target event in $96.9$\% of Wave~1 iterations and $98.5$\% of Wave~2 iterations on the ``in set'' metric, and in $95.4$\%/$95.1$\% of iterations on the ``top-10'' metric. The remaining gap reflects two sources of error: the generation model occasionally truncates its tagged id list before reaching the target, and the retriever's similarity cutoff and date pre-filter can drop the target when the query phrasing departs from the indexed event's tokens. Because the hybrid system pools the two retrieval streams, it inherits the filter's perfect baseline.

\clearpage

\section{Balance Tests}\label{app:balance}

Table~\ref{tab:balance} reports means by treatment for pre-treatment covariates measured in the post-task demographic survey. The treatments are well balanced on age, gender, income, NYC proximity, residence in New Jersey, and quiz performance. The only pairwise difference significant at the 5\% level is on the share of participants holding at least a four-year college degree, which is higher in \textit{Chat} (37.8\%) than in \textit{Filter} (24.0\%). All regression specifications in the main text include education as a control.
 
\begin{table}[h]
\centering
\small
\caption{Balance tests across treatments (analysis sample, $N=362$).}
\label{tab:balance}
\begin{tabular}{lrrrr}
\hline
Variable & \textit{Filter}$^a$ & \textit{Hybrid}$^b$ & \textit{Chat}$^c$ & $N$ \\
\hline
Age (years)                                  & 38.73            & 37.05         & 37.78           & 362 \\
Share female                                 & 0.606            & 0.510         & 0.494           & 362 \\
Share with Bachelor's degree or higher       & 0.240$^{c}$      & 0.265         & 0.378$^{a}$     & 362 \\
Share with graduate degree                   & 0.096            & 0.108         & 0.141           & 362 \\
Income category (1--6)                       & 2.442            & 2.618         & 2.385           & 362 \\
NYC proximity (1 = closest)                  & 2.269            & 2.157         & 2.353           & 362 \\
Share NJ resident                            & 0.202            & 0.206         & 0.231           & 362 \\
Total quiz errors                            & 0.635            & 0.696         & 0.538           & 362 \\
\hline
$N$ per treatment                            & 104              & 102           & 156             &     \\
\hline
\end{tabular}

\vspace{3pt}

\begin{minipage}{0.64\linewidth}
\footnotesize\linespread{0.9}\selectfont
\textit{Notes:} Cell entries are means by treatment. A superscript on a cell indicates that the cell differs from the corresponding-letter treatment at $p < 0.05$ in a two-sided $t$-test (a = \textit{Filter}, b = \textit{Hybrid}, c = \textit{Chat}). Income is coded on a six-point scale from $<\$25{,}000$ to $>\$150{,}000$. NYC proximity is coded on a four-point scale where 1 = ``in New York City'' and 4 = ``outside the New York metropolitan area.''\par
\end{minipage}
\end{table}
 
\clearpage

\section{Pilot Studies (Detail)}\label{app:pilots}

The final design was developed through two pilot studies. Both were run on CloudResearch with NY/NJ residents and used the same scraped Ticketmaster catalog as the main study.

\textbf{Pilot 1 (February 27, 2026; $N = 40$, \textit{Filter} $= 21$, \textit{Chat} $= 19$).} The first pilot served three purposes. First, it allowed us to validate the BDM elicitation scale. The median switching point in the pilot was \$25 and fewer than 10\% of participants hit the upper bound, indicating that a \$0--\$50 range was wide enough to avoid systematic ceiling effects without producing a long, sparsely populated upper tail. Second, the pilot stress-tested the conversational interface. We logged every user message and AI response, identified five recurring failure modes, and tightened the RAG service to address each. The five failure modes were: (i) broad-topic queries (e.g., \emph{``indie''}) returning padded irrelevant results; (ii) brittle handling of negation and exclusion (e.g., \emph{``not including museums or theater''}); (iii) no support for borough-level constraints; (iv) weak handling of team and venue abbreviations (e.g., \emph{``NYK''} for the Knicks); and (v) occasional confident misstatements when the retrieved candidates did not actually match the request (e.g., the model claiming to have found soccer events when the candidates were basketball games). The post-pilot fixes included a stricter relevance cutoff in the retrieval step, an explicit team-and-venue alias dictionary, a borough pre-filter, and a strengthened system prompt instructing the model to acknowledge near-empty result sets rather than mischaracterize them. Third, the pilot allowed us to assess the structured filter. Participant feedback (e.g., requests for a borough filter and a calendar-style date selector) led us to expand the filter panel for the main study.

\textbf{Pilot 2 (March 19, 2026; $N = 32$ completers, \textit{Filter} $= 8$, \textit{Chat} $= 11$, \textit{Hybrid} $= 13$).} The second pilot was the first to include the new \textit{Hybrid} interface. Sample sizes were intentionally small, since this pilot was designed to validate the post-pilot-1 fixes, confirm that the combined interface was usable, and provide a final read on session timing, comprehension-quiz difficulty, and exclusion-rule calibration. Pilot 2 confirmed that the AI fixes deployed after Pilot 1 had eliminated the most salient failure modes, and that participants in \textit{Hybrid} were able to use both modalities without confusion.

\clearpage

\section{Experimental Interface and Instructions}\label{app:materials}

This section reproduces the participant-facing text and the interface used in the \textit{Chat} treatment. The text used in the \textit{Filter} and \textit{Hybrid} treatments is identical in the welcome and consent pages and differs only in the treatment-specific instructions and comprehension quiz, which are adapted to describe the relevant interface.

\subsection*{Welcome and Consent}

After clearing a brief bot-detection screen and a New York / New Jersey residency screen, participants are shown the following welcome and consent page (\textit{instructions\_0}):

\begin{verbatimbox}\small
Welcome to this decision-making study. The study takes around 15--20 minutes to complete. Please note that this study \textbf{cannot be completed on the Safari browser}.

Please pay close attention to the instructions. To ensure that you understand the instructions we will ask you several questions as we go along. You will only be allowed to proceed to the task if you can answer those questions.

If you complete this study you will receive a payment of \$4. In addition, one of the study participants will be selected using a random draw at the end of this session (which consists of 20 participants). That person will receive an additional bonus, which will be explained later.

To continue please review the consent form below.

\textbf{Informed Consent Information.} \textbf{Title:} Decision-Making Study.

You are being invited to participate in a research study examining decision-making in consumer search. Your participation will involve performing a search task over several experiential products, which will take approximately 15--20 minutes. Participation is voluntary, your responses will be kept confidential and de-identified, and you may withdraw at any time without penalty. This study involves minimal risk and has been determined exempt from ongoing IRB review. You will receive a payment of \$4. In addition, one person in this session will be selected at random and will receive the experience they select, or a monetary amount of up to \$50. By clicking ``Next'' below, you agree to participate in this study. For questions, contact \textit{(author email redacted for review)} or the IRB at \textit{(institution redacted for review)}.

If you are 18 years of age or older, understand the statements above, and freely consent to participate in the study, click on the ``I Agree'' button to begin the experiment.
\end{verbatimbox}

\subsection*{Task Instructions (\textit{Chat})}

After consent, participants in the \textit{Chat} treatment see the following task instructions (\textit{A01\_Instructions}):

\begin{verbatimbox}\small
\textbf{Overview.} In this study, you will search through a collection of events taking place in \textbf{New York City in May and June 2026}. These events include concerts, theater, sports, and more.

\textbf{Your Task.} The number of events is very large and you will not be able to explore all of them. Your goal is to find and select your \textbf{favorite event} from this collection. You can search as many times as you want. Once you find an event you like, select it and proceed.

\textbf{What Happens After You Select an Event?} After completing the study, you will receive:
\begin{itemize}\itemsep 0pt
\item A \textbf{guaranteed payment of \$4.00}.
\item Entry into a \textbf{lottery} where you have a chance to win either a ticket to the event you selected, or some amount of money (up to \$50).
\end{itemize}

Each session consists of 20 participants. One winner will be drawn at the end of the session. If you win the event option, we will provide the cheapest available ticket for your selected event.

\textbf{This is real!} The lottery winner will actually receive either the event ticket or money. This is not hypothetical.

\textbf{Important Notes.}
\begin{itemize}\itemsep 0pt
\item Think carefully about which event you would most want to attend---you might actually win a ticket to that event.
\item Take your time exploring the options before making your final selection.
\end{itemize}
\end{verbatimbox}

\subsection*{Comprehension Quiz}

Participants then complete a three-question comprehension quiz (\textit{P01\_Instructions}). Participants who answer four or more questions incorrectly across all three items are excluded from analysis (\textsection\ref{sec:meth-overview}). Correct answers are marked in bold.

\begin{verbatimbox}\small
\textbf{Q1: I will be able to explore all of the available events. True or False?}
\begin{enumerate}\itemsep 0pt
\item[a)] True.
\item[b)] False, I will only get three events to choose from.
\item[\textbf{c)}] \textbf{False, the number of available events is very large and I will only see a subset of them.}
\item[d)] False, I will only get ten events to choose from.
\end{enumerate}

\textbf{Q2: The task is entirely hypothetical and I will only receive a fixed payment at the end of the experiment. True or False?}
\begin{enumerate}\itemsep 0pt
\item[a)] True.
\item[b)] False, I will also receive the event I selected.
\item[c)] False, I will also be entered into a drawing to receive some amount of money (up to \$50).
\item[\textbf{d)}] \textbf{False, I will also be entered into a drawing to receive either the event I selected or some amount of money (up to \$50).}
\end{enumerate}

\textbf{Q3: I can only perform the search once. True or False?}
\begin{enumerate}\itemsep 0pt
\item[a)] True.
\item[b)] False, I will be allowed to perform the search twice.
\item[\textbf{c)}] \textbf{False, I will be allowed to go back to the initial search as much as I want.}
\end{enumerate}
\end{verbatimbox}

\subsection*{Subsequent Pages}

\begin{figure}[h]
\caption{Valuation elicitation screen (multiple-price-list implementation of BDM).}
\label{fig:mpl}
\centering
\includegraphics[width=0.67\linewidth]{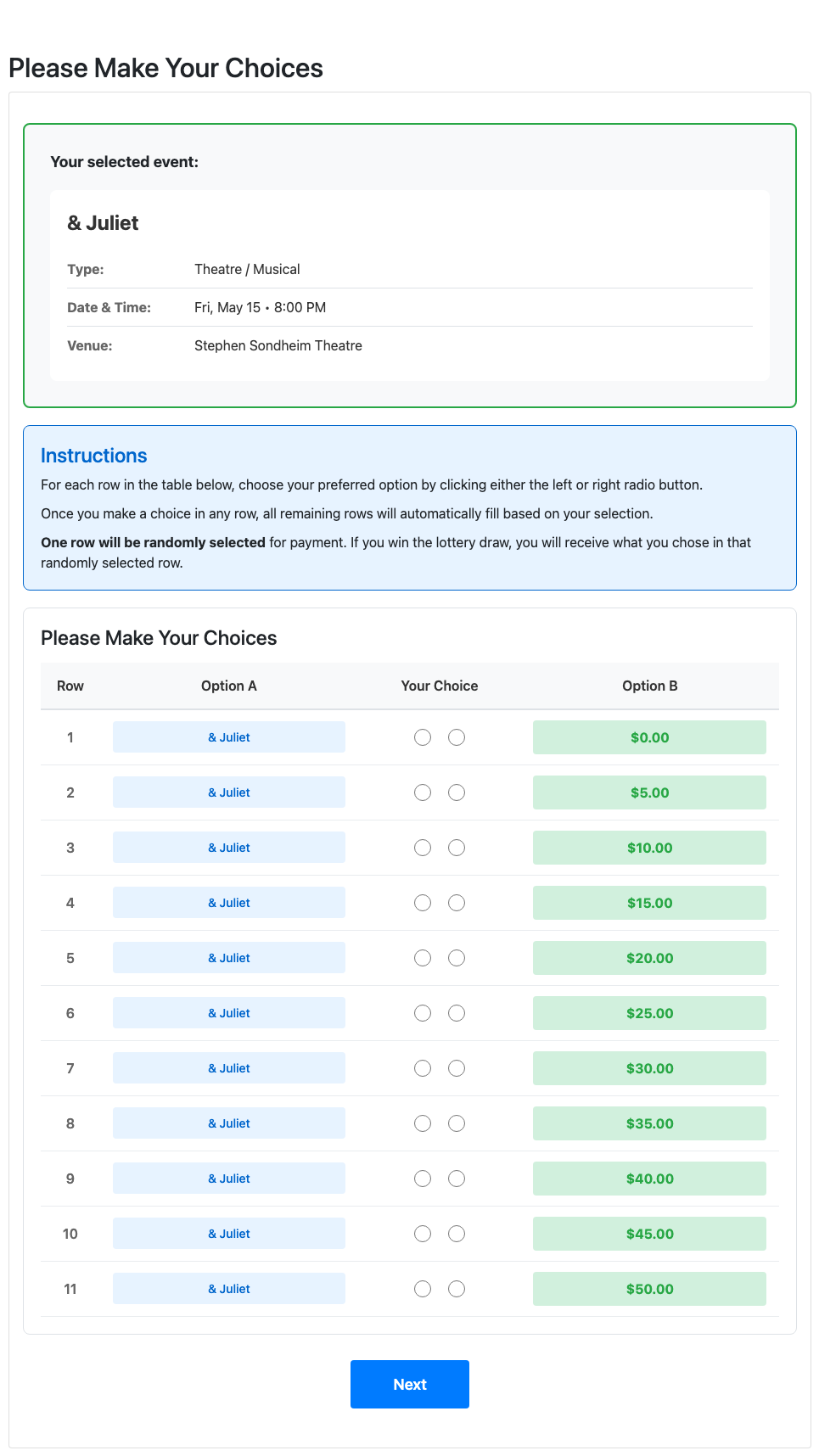}
\end{figure}

After the quiz, participants proceed to the search task itself (Figure~\ref{fig:screen-genai}), then to a confirmation page where they review the selected event, and then to the BDM valuation elicitation. The valuation screen is shown in Figure~\ref{fig:mpl}: each participant sees the event they just selected at the top, followed by an 11-row multiple price list comparing that event against cash amounts ranging from \$0 to \$50 in \$5 increments. After the elicitation, participants are shown a lottery-outcome page, a post-task survey collecting demographics and self-reported satisfaction, and a completion page that returns the CloudResearch validation code.

\clearpage

\section{AI Model Performance and Search Intent (Detail)}\label{app:model-intent}
This section supplements \textsection\ref{sec:res-model-intent-model}: \textsection\ref{app:modelreg} examines robustness to the AI model, and \textsection\ref{app:intent} reports the treatment\,$\times$\,search-intent interaction models.

\subsection{Robustness to the AI Model}\label{app:modelreg}
The bar charts in Figure~\ref{fig:model_comparison} compared the three AI models within the \textit{Chat} arm using means and rank tests. Table~\ref{tab:model_reg} reports the regression analogue: each outcome is regressed on indicators for the two frontier models (relative to the legacy \texttt{gpt-4.1-mini}) and the same controls as the main specification, estimated on the \textit{Chat} subsample ($N=156$); Figure~\ref{fig:model_comparison} in the main text plots the three pairwise model contrasts for each outcome. No model coefficient is significant, the model indicators are jointly insignificant for every outcome, and all pairwise contrasts include zero.

\begin{table}[htbp]\centering
\caption{Outcomes by AI model (within the \textit{Chat} arm)}\label{tab:model_reg}
\small
\sbox{\regtablebox}{%
\begin{tabular}{l *{4}{>{\centering\arraybackslash}m{2.4cm}}}
\toprule
\multicolumn{1}{r}{Dep.\ var.:} & \textit{Valuation} & \textit{Outcome Satisfaction} & \textit{Search Time} & \textit{Process Satisfaction} \\
\cmidrule(lr){2-2}\cmidrule(lr){3-3}\cmidrule(lr){4-4}\cmidrule(lr){5-5}
\multicolumn{1}{r}{Estimator:} & Interval & OLS & OLS & OLS \\
 & (1) & (2) & (3) & (4) \\
\midrule
\texttt{gpt-4.1-mini} & Omitted & Omitted & Omitted & Omitted \\
\addlinespace
\texttt{gpt-5.2} & $-$4.896 & $-$0.104 & $-$28.898 & 0.020 \\
 & (5.395) & (0.363) & (36.951) & (0.399) \\
\addlinespace
\texttt{gpt-5.4} & $-$5.135 & $-$0.104 & $-$40.939 & 0.307 \\
 & (5.414) & (0.364) & (32.209) & (0.358) \\
\midrule
Controls & Yes & Yes & Yes & Yes \\
Observations & 156 & 156 & 156 & 156 \\
$R^2$ (adj.; $^{\dagger}$pseudo) & 0.026$^{\dagger}$ & -0.022 & 0.089 & -0.011 \\
Joint $p$ (model $=0$) & 0.56 & 0.95 & 0.44 & 0.59 \\
\midrule
\texttt{gpt-5.4} $-$ \texttt{gpt-5.2} & $-$0.238 & 0.001 & $-$12.041 & 0.287 \\
 & (5.419) & (0.349) & (32.459) & (0.340) \\
\bottomrule
\end{tabular}}%
\usebox{\regtablebox}\par
\vspace{3pt}
\begin{minipage}{\wd\regtablebox}
\footnotesize\linespread{0.9}\selectfont
\textit{Notes:} Estimated within the \textit{Chat} arm ($N=156$), with the AI model randomized across participants. Omitted model is \texttt{gpt-4.1-mini}. Standard errors in parentheses (HC3 robust for OLS; model-based for the interval regression). Column~1 is an interval regression (each price-list response enters as the interval it revealed; the top category is censored above \$50); columns 2--4 are OLS. Controls match Table~\ref{tab:reg_main}. $^{***}p<0.01$, $^{**}p<0.05$, $^{*}p<0.1$.\par
\end{minipage}
\end{table}

\clearpage

\subsection{Treatment Effects by Search Intent}\label{app:intent}
Table~\ref{tab:intent_appx} reports the treatment\,$\times$\,search-intent interaction models for valuation and the two satisfaction measures (\textsection\ref{sec:res-model-intent-intent}). Column~1 uses the same interval-regression specification as the valuation column of Table~\ref{tab:reg_main}; columns~2--3 are OLS. The upper block gives the main effects and the interaction coefficients; the lower block gives the three pairwise treatment contrasts at each of the three intent levels.
\begin{table}[h!]\centering
\caption{Treatment $\times$ search-intent interaction: valuation and satisfaction}\label{tab:intent_appx}
\footnotesize
\renewcommand{\arraystretch}{0.88}%
\sbox{\regtablebox}{%
\begin{tabular}{l *{3}{>{\centering\arraybackslash}m{2.6cm}}}
\toprule
\multicolumn{1}{r}{Dep.\ var.:} & \textit{Valuation} & \textit{Outcome Satisfaction} & \textit{Process Satisfaction} \\
\cmidrule(lr){2-2}\cmidrule(lr){3-3}\cmidrule(lr){4-4}
\multicolumn{1}{r}{Estimator:} & Interval & OLS & OLS \\
 & (1) & (2) & (3) \\
\midrule
\textit{Chat} & $-$15.894$^{***}$ & $-$0.685$^{**}$ & $-$0.487 \\
 & (5.776) & (0.308) & (0.325) \\
\addlinespace[3pt]
\textit{Hybrid} & 5.248 & 0.313 & $-$0.121 \\
 & (6.593) & (0.274) & (0.343) \\
\addlinespace[3pt]
Specific & $-$3.992 & 0.935$^{***}$ & 1.059$^{***}$ \\
 & (7.624) & (0.271) & (0.349) \\
\addlinespace[3pt]
In-between & $-$0.290 & $-$0.026 & 0.214 \\
 & (5.980) & (0.275) & (0.305) \\
\addlinespace[3pt]
\textit{Chat} $\times$ Specific & 24.243$^{**}$ & $-$0.326 & $-$0.604 \\
 & (10.087) & (0.506) & (0.565) \\
\addlinespace[3pt]
\textit{Chat} $\times$ In-between & 17.225$^{**}$ & 0.297 & $-$0.164 \\
 & (7.617) & (0.400) & (0.424) \\
\addlinespace[3pt]
\textit{Hybrid} $\times$ Specific & $-$2.336 & $-$0.612 & $-$0.251 \\
 & (10.690) & (0.456) & (0.544) \\
\addlinespace[3pt]
\textit{Hybrid} $\times$ In-between & $-$2.221 & $-$0.049 & $-$0.180 \\
 & (8.594) & (0.381) & (0.454) \\
\addlinespace[3pt]
\midrule
Demographic Controls & Yes & Yes & Yes \\
Observations & 362 & 362 & 362 \\
$R^2$ (adj.; $^{\dagger}$pseudo) & 0.037$^{\dagger}$ & 0.086 & 0.071 \\
\midrule
\multicolumn{4}{l}{\textit{Pairwise treatment contrasts:}} \\ \addlinespace[3pt]
\quad Open-ended intent: \textit{Chat} $-$ \textit{Filter} & $-$15.89$^{***}$ & $-$0.68$^{**}$ & $-$0.49 \\
 & (5.78) & (0.31) & (0.33) \\
\addlinespace[3pt]
\quad Open-ended intent: \textit{Hybrid} $-$ \textit{Filter} & 5.25 & 0.31 & $-$0.12 \\
 & (6.59) & (0.27) & (0.34) \\
\addlinespace[3pt]
\quad Open-ended intent: \textit{Hybrid} $-$ \textit{Chat} & 21.14$^{***}$ & 1.00$^{***}$ & 0.37 \\
 & (5.94) & (0.30) & (0.35) \\
\addlinespace[3pt]
\quad In-between intent: \textit{Chat} $-$ \textit{Filter} & 1.33 & $-$0.39 & $-$0.65$^{**}$ \\
 & (5.03) & (0.27) & (0.28) \\
\addlinespace[3pt]
\quad In-between intent: \textit{Hybrid} $-$ \textit{Filter} & 3.03 & 0.26 & $-$0.30 \\
 & (5.51) & (0.26) & (0.30) \\
\addlinespace[3pt]
\quad In-between intent: \textit{Hybrid} $-$ \textit{Chat} & 1.70 & 0.65$^{**}$ & 0.35 \\
 & (5.04) & (0.26) & (0.30) \\
\addlinespace[3pt]
\quad Specific intent: \textit{Chat} $-$ \textit{Filter} & 8.35 & $-$1.01$^{**}$ & $-$1.09$^{**}$ \\
 & (8.36) & (0.41) & (0.47) \\
\addlinespace[3pt]
\quad Specific intent: \textit{Hybrid} $-$ \textit{Filter} & 2.91 & $-$0.30 & $-$0.37 \\
 & (8.42) & (0.37) & (0.41) \\
\addlinespace[3pt]
\quad Specific intent: \textit{Hybrid} $-$ \textit{Chat} & $-$5.44 & 0.71 & 0.72 \\
 & (8.01) & (0.48) & (0.49) \\
\addlinespace[3pt]
\bottomrule
\end{tabular}}%
\usebox{\regtablebox}\par
\vspace{3pt}
\begin{minipage}{\wd\regtablebox}
\footnotesize\linespread{0.9}\selectfont
\textit{Notes:} Column 1 is an interval regression (each price-list response enters as the interval it revealed; the top category is censored above \$50) with model-based standard errors; columns 2--3 are OLS with HC3 robust standard errors. Standard errors in parentheses. Controls match Table~\ref{tab:reg_main}. The lower block reports the three pairwise treatment contrasts at each search-intent level. $^{***}p<0.01$, $^{**}p<0.05$, $^{*}p<0.1$.\par
\end{minipage}
\end{table}

\clearpage

\section{Mechanisms}\label{app:mech}

\subsection{Variable Definitions}
\label{app:mech_variables}
\subsubsection*{Interaction logs}
Each participant's session is captured as a chronological log of every action taken inside the search interface. We use three running examples throughout this section to illustrate the variables we construct from these logs. Figure~\ref{fig:filterexample} shows a \textit{Filter} session, where $t$ is the time in seconds from the start of the session. The participant first narrows the catalog to Music events using the genre drop-down, refines by typing \texttt{radio city} (potentially looking for events at Radio City Music Hall), adds a date constraint of 2026-05-07, then types \texttt{mitski} to look up a specific performer, before pivoting to Arts \& Theater and searching for \texttt{hamilton}, the Broadway show.

\begin{figure}[htbp]
\vspace{-0.25cm}
	\centering
	\scriptsize
	\caption{Example of a \textit{Filter} session}
	\vspace{0cm}
	\label{fig:filterexample}
	\begin{minipage}{\textwidth}
		\setlength{\topsep}{1pt}\setlength{\partopsep}{0pt}%
		\begin{verbatim}
			t=0    update_filters   genre="Music"
			t=21   text_search      genre="Music"           text="radio city"
			t=31   update_filters   genre="Music"           date="2026-05-07"
			t=36   text_search      genre="Music"           text="mitski"        date="2026-05-07"
			t=53   update_filters   genre="Arts & Theater"                       date="2026-05-07"
			t=72   text_search      genre="Arts & Theater"  text="hamilton"      date="2026-05-07"
		\end{verbatim}
		\vspace{-0.85\baselineskip}
	\end{minipage}
\vspace{-0.15cm}
\end{figure}
\vspace{-0.15cm}

Figure~\ref{fig:genaiexample} shows a \textit{Chat} session. The participant first asks for \texttt{Concert}, then narrows by genre and date with the request \emph{Rock or alternative concerts in April or May 2026}, then sends a near-duplicate refinement restricted to May.

\begin{figure}[htbp]
\vspace{-0.35cm}
	\centering
	\scriptsize
	\caption{Example of a \textit{Chat} session}
	\vspace{0cm}
	\label{fig:genaiexample}
	\begin{minipage}{\textwidth}
	\setlength{\topsep}{1pt}\setlength{\partopsep}{0pt}%
	\begin{verbatim}
		t=0    chat_message     role="user"   text="Concert"
		t=1    chat_response    ...
		t=34   chat_message     role="user"   text="Rock or alternative concerts in April or May 2026"
		t=39   chat_response    ...
		t=113  chat_message     role="user"   text="Rock or alternative concerts in May 2026"
	\end{verbatim}
	\vspace{-0.85\baselineskip}
	\end{minipage}
\vspace{-0.25cm}
\end{figure}

\vspace{-0.25cm}

Figure~\ref{fig:hybridexample} shows a \textit{Hybrid} session, which combines the filter and chat interfaces. The participant first narrows to Arts \& Theater via the drop-down, uses the chat to ask about \texttt{painting} and \texttt{drawing} (neither of which corresponds to any catalog event), then returns to the drop-down to pivot to Music and back to Arts \& Theater.

\begin{figure}[htbp]
\vspace{-0.15cm}
	\centering
	\scriptsize
	\caption{Example of a \textit{Hybrid} session}
	\vspace{0cm}
	\label{fig:hybridexample}
	\begin{minipage}{\textwidth}
\setlength{\topsep}{1pt}\setlength{\partopsep}{0pt}%
\begin{verbatim}
	t=0    filter_change    genre="Arts & Theater"
	t=59   chat_message     role="user"   text="painting"
	t=101  chat_message     role="user"   text="drawing"
	t=138  filter_change    genre="Music"
	t=237  filter_change    genre="Arts & Theater"
\end{verbatim}
\vspace{-0.85\baselineskip}
\end{minipage}
\vspace{-0.15cm}
\end{figure}
\vspace{-0.25cm}

\subsubsection*{Constructing the variables}
We construct three counts: \texttt{search breadth}, the number of distinct segments searched (\textit{Music}, \textit{Sports}, or \textit{Arts \& Theater}); \texttt{granularity}, the number of distinct subgenres searched; and \texttt{out-of-catalog}, the number of distinct out-of-catalog searches. We label each search action along three dimensions: \textbf{T} if it initiates a search in a new segment; \textbf{S} if it initiates a search in a new subgenre (e.g.\ \textit{Rock} or \textit{MLB}); and \textbf{O} if it refers to an event, venue, or a concept that matches no catalog entry. We read segment labels directly from the drop-down in \textit{Filter} and \textit{Hybrid}, but infer them from free text in \textit{Chat}; we infer subgenre and out-of-catalog labels from free text in all three arms.

The distinction between a subgenre concept (\textbf{S}) and an out-of-catalog concept (\textbf{O}) is determined by looking the input up against the event catalog. We check each typed \textit{Filter} query or chat-message phrase against four fields:

\begin{itemize}
	\setlength\itemsep{1pt}
	\item[(i)] the 725 distinct \texttt{event\_name}s (e.g.\ \emph{Hamilton}, \emph{Mitski}, \emph{The Lion King}); a match resolves directly to the event's (segment, subgenre) pair;
	\item[(ii)] the 136 distinct venue names (e.g.\ \emph{Radio City Music Hall}, \emph{Citi Field}); for venues hosting events across multiple subgenres, the lookup returns the modal subgenre, with an alphabetical tie-break;
	\item[(iii)] the 107 subgenres (e.g.\ \emph{Musical}, \emph{Rock}, \emph{Jazz}); and
	\item[(iv)] the three top-level segments (\emph{Arts \& Theater}, \emph{Music}, \emph{Sports}).
\end{itemize}

A fuzzy match at any of (i)--(iii) resolves the input to a subgenre and yields an \textbf{S} label; an input that fails all four lookups yields an \textbf{O} label and is recorded as out-of-catalog.

Applying these rules to the running examples clarifies the construction. In the \textit{Filter} session, \texttt{radio city} matches no event name, does not resolve through the venue lookup as a standalone string, and matches no subgenre or segment, so it is out-of-catalog; \texttt{mitski} matches the event name \emph{Mitski} at (i), resolving to (\emph{Music}, \emph{Alternative}); and \texttt{hamilton} matches \emph{Hamilton} at (i), resolving to (\emph{Arts \& Theater}, \emph{Musical}). The session therefore yields $\texttt{search breadth}=2$ (the drop-down selections \emph{Music} and \emph{Arts \& Theater}), $\texttt{granularity}=2$ (\emph{Alternative} from \texttt{mitski} and \emph{Musical} from \texttt{hamilton}), and $\texttt{out-of-catalog}=1$ (\texttt{radio city}). In the \textit{Chat} session, \texttt{Concert} matches the \emph{Music} segment at (iv) and contributes a segment, while \texttt{Rock} and \texttt{Alternative} match at (iii) and contribute subgenres. In the \textit{Hybrid} session, the two drop-down selections give $\texttt{search breadth}=2$ (\emph{Arts \& Theater}, \emph{Music}), the two chat messages give $\texttt{out-of-catalog}=2$ (\texttt{painting}, \texttt{drawing}), and no subgenre-level content is added ($\texttt{granularity}=0$).

In summary, each count tallies the distinct concepts a participant expressed along one catalog dimension over the session: \texttt{search breadth} counts distinct \textbf{T} labels, \texttt{granularity} distinct \textbf{S} labels, and \texttt{out-of-catalog} distinct \textbf{O} labels. Note, however, that \texttt{out-of-catalog} intent is only partially observed under \textit{Filter} and \textit{Hybrid}, because we observe it only when a participant types a missing item into the search box or chat. A participant looking for ``Taylor Swift'' under \textit{Filter} may scan the \textit{Music} drop-down, find nothing matching, and settle for an alternative without ever typing the artist's name. The \textit{Filter} and \textit{Hybrid} counts of \texttt{out-of-catalog} are therefore lower bounds. 




\subsection{Search-Process Measures: Full Set}\label{app:process-full}

Table~\ref{tab:process} reports the full set of search-process measures by interface, together with tests of treatment differences and each measure's raw correlation with valuation. Panels~A1--A2 show how participants use different features of each interface, and how the results returned by each interface differ. The differences between the interfaces align with intuition: in the \textit{Chat} treatment users make a larger number of queries (since they lack the filter option), but the number of events, segments and subgenres displayed to them is smaller. However, with the exception of the use of the filter menu (which is nearly perfectly correlated with treatment assignment), none of these measures is predictive of event valuation.\footnote{We focus on valuation because it is our only incentivized measure. However, similar results hold if we examine other dependent variables (our self-reported satisfaction measures).}

\begin{table}[htbp]
\centering
\caption{Search process measures by interface}
\label{tab:process}
\small
\begin{tabular}{c l ccc c c}
\toprule
 & & \textit{Filter} & \textit{Chat} & \textit{Hybrid} & $p$ & Corr.\ with \\
 & & ($N=104$) & ($N=156$) & ($N=102$) & (across arms) & valuation \\
\midrule
\multirow{12}{*}{\rotatebox{90}{\shortstack{\textit{\textbf{General process}}\\\textit{\textbf{Measures}}}}}
 & \multicolumn{6}{l}{\textit{Panel A1: User Behavior}} \\
  & \quad Used filter menu & 97\% & 0\% & 96\% & $<0.001$ & $+0.09^{*}$ \\
   & \quad Number of filter changes  & 4.61 & 0.00 & 3.54 & $<0.001$ & $+0.04$ \\
 & \quad Used text/chat queries & 55\% & 100\% & 49\% & $<0.001$ & $-0.04$ \\
 & \quad Number of text or chat queries  & 1.72 & 3.91 & 1.20 & $<0.001$ & $-0.03$ \\
 & \quad Number of words & 1.98 & 18.15 & 3.44 & $<0.001$ & $-0.09$ \\
 & \quad Reset their search & 13\% & 10\% & 11\% & $0.577$ & $+0.03$ \\
 \addlinespace
  & \multicolumn{6}{l}{\textit{Panel A2: Results Returned}} \\
 & \quad Events displayed  & 36.4 & 25.0 & 32.2 & $0.002$ & $+0.04$ \\
 & \quad Distinct segments displayed & 2.47 & 2.10 & 2.33 & $<0.001$ & $+0.02$ \\
 & \quad Distinct subgenres displayed & 14.49 & 7.69 & 12.95 & $<0.001$ & $+0.04$ \\
\midrule
\multirow{6}{*}{\rotatebox{90}{\shortstack{\textit{\textbf{\textit{Echo chamber}}}\\\textit{\textbf{mechanism}}}}}
 & \multicolumn{6}{l}{\textit{Panel B1: Breadth/Granularity of search}} \\
 & \quad Distinct segments searched & 1.87 & 1.10 & 1.62 & $<0.001$ & $+0.13^{**}$ \\
 & \quad Distinct subgenres searched & 0.59 & 1.02 & 0.48 & $<0.001$ & $+0.06$ \\
 \addlinespace 
 & \multicolumn{6}{l}{\textit{Panel B2: Out-of-catalog search}} \\
 & \quad Any out-of-catalog query & 29\% & 49\% & 19\% & $<0.001$ & $-0.08$ \\
 & \quad Out-of-catalog queries  & 0.60 & 1.17 & 0.36 & $<0.001$ & $-0.09^{*}$ \\
\midrule
\multirow{12}{*}{\rotatebox{90}{\shortstack{\textit{\textbf{\textit{Prompting fatigue}}}\\\textit{\textbf{mechanism}}}}}
 & \multicolumn{6}{l}{\textit{Panel C1: Time allocation}} \\
 & \quad Session time (Seconds) & 165 & 214 & 175 & $<0.001$ & $-0.04$ \\
 & \quad Composing queries (Seconds) & 3.8 & 33.2 & 6.1 & $<0.001$ & $-0.08$ \\
 & \quad Waiting on AI (Seconds) & 0.0 & 40.0 & 11.7 & $<0.001$ & $-0.09^{*}$ \\
 & \quad Inspecting results (Seconds) & 161 & 141 & 157 & $0.559$ & $-0.01$ \\
\addlinespace
 & \quad [Composing queries]/[Session Time] & 3\% & 15\% & 3\% & $<0.001$ & $-0.09^{*}$ \\
 & \quad [Waiting on AI]/[Session Time] & 0\% & 21\% & 6\% & $<0.001$ & $-0.07$ \\
 & \quad [Inspecting results]/[Session Time] & 97\% & 63\% & 91\% & $<0.001$ & $+0.09^{*}$ \\
 \addlinespace 
 
 & \multicolumn{6}{l}{\textit{Panel C2: Position of the chosen event}} \\
 & \quad Stayed on first 10 results & 49\% & 63\% & 53\% & $0.068$ & $-0.14^{***}$ \\
 & \quad Stayed on first 3 results & 31\% & 46\% & 34\% & $0.036$ & $-0.11^{**}$ \\
 & \quad Stayed on first result & 12\% & 23\% & 14\% & $0.030$ & $-0.06$ \\
\bottomrule
\end{tabular}
\vspace{3pt}
\begin{minipage}{0.95\linewidth}
\footnotesize\linespread{0.9}\selectfont \textit{Notes:} Cells are treatment arm means. \emph{Segments} are the three top-level categories (Sports, music, etc.). \emph{Subgenres} are finer categories (e.g., Jazz or Blues within Music); $p$-values are based on tests of equality across treatment arms ($\chi^2$ for shares, Kruskal--Wallis for means); Corr. shows pairwise Pearson correlation coefficients between each process measure and valuation. In the time-allocation rows (Panel~C1), waiting is the measured AI response time, composing is typed characters divided by a typing rate of 3 characters per second, and inspecting is the remainder of the session. In Panel~C2, ``stayed on first $k$'' means the chosen event was among the first $k$ positions of the results list it was chosen from. $^{*}\,p<0.10$, $^{**}\,p<0.05$, $^{***}\,p<0.01$. \par
\end{minipage}
\end{table}

\subsection{Search-Process Margins: Full Specification}\label{app:echo-spec}
Table~\ref{tab:echo_spec} reports the interval regressions underlying Figure~\ref{fig:echo_wtp}. For each process measure we estimate a valuation interval regression on the measure in its original units, pooled across treatment arms, with the same demographic controls as the main valuation regressions. The table reports the latent-scale coefficient and the implied average marginal effect on expected valuation; the latter are the effects annotated in Figure~\ref{fig:echo_wtp}.

\begin{table}[htbp]\centering
\caption{Regressions underlying Figure~\ref{fig:echo_wtp}}\label{tab:echo_spec}
\small
\sbox{\regtablebox}{%
\begin{tabular}{l *{2}{>{\centering\arraybackslash}m{3.0cm}}}
\toprule
\multicolumn{1}{r}{Dep.\ var.:} & \multicolumn{2}{c}{\textit{Valuation}} \\
\cmidrule(lr){2-3}
\multicolumn{1}{r}{Process measure:} & \textit{Segments} & \textit{Out-of-catalog} \\
\cmidrule(lr){2-2}\cmidrule(lr){3-3}
 & (1) & (2) \\
\midrule
Process measure (latent) & 3.50$^{**}$ & $-$2.28$^{**}$ \\
 & (1.76) & (1.04) \\
\midrule
AME on $E[\text{valuation}]$ & 2.05$^{**}$ & $-$1.34$^{**}$ \\
 & (1.03) & (0.61) \\
\midrule
Demographic controls & Yes & Yes \\
Observations & 362 & 362 \\
\bottomrule
\end{tabular}}%
\usebox{\regtablebox}\par
\vspace{3pt}
\begin{minipage}{\wd\regtablebox}
\footnotesize\linespread{0.9}\selectfont
\textit{Notes:} Interval regressions of valuation (each price-list response enters as the interval it revealed; the top category is censored above \$50) on each process measure in original units, pooled across treatment arms (no treatment terms). Standard errors in parentheses (model-based). AME denotes the latent coefficient scaled by the mean probability of an interior observation; the AMEs are the effects annotated in Figure~\ref{fig:echo_wtp}. $^{***}p<0.01$, $^{**}p<0.05$, $^{*}p<0.1$.\par
\end{minipage}
\end{table}

\subsection{\textit{Prompting Fatigue}: Underlying Regressions}\label{app:attention}

Table~\ref{tab:attention_spec} reports the regressions displayed in Figure~\ref{fig:attention_panels}: a linear probability model of choosing an event from the first ten positions of the results list on the \textit{CW share} of the session (the slope annotated in panel b), and an interval regression of valuation on an indicator for choosing beyond the first ten positions (the effect annotated in panel c). Both specifications are pooled across treatment arms and include the demographic controls of Table~\ref{tab:reg_main}.

\begin{table}[htbp]\centering
\caption{Regressions underlying Figure~\ref{fig:attention_panels}}\label{tab:attention_spec}
\small
\sbox{\regtablebox}{%
\begin{tabular}{l *{2}{>{\centering\arraybackslash}m{3.0cm}}}
\toprule
\multicolumn{1}{r}{Dep.\ var.:} & \textit{Chose from top 10} & \textit{Valuation} \\
\cmidrule(lr){2-2}\cmidrule(lr){3-3}
 & (1) & (2) \\
\midrule
\textit{CW share} of session & 0.58$^{***}$ &  \\
 & (0.11) &  \\
\addlinespace
Chose beyond top 10 (latent) &  & 7.17$^{**}$ \\
 &  & (2.88) \\
\midrule
AME on $E[\text{valuation}]$ &  & 4.21$^{**}$ \\
 &  & (1.69) \\
\midrule
Demographic controls & Yes & Yes \\
Observations & 362 & 362 \\
\bottomrule
\end{tabular}}%
\usebox{\regtablebox}\par
\vspace{3pt}
\begin{minipage}{\wd\regtablebox}
\footnotesize\linespread{0.9}\selectfont
\textit{Notes:} Pooled specifications without treatment terms. Column (1): linear probability model of choosing an event from the first ten positions on the \textit{CW share} of the session (HC1 standard errors); this slope is annotated in Figure~\ref{fig:attention_panels}(b). Column (2): interval regression of valuation (each price-list response enters as its revealed interval; the top category is censored above \$50) on an indicator for choosing beyond the first ten positions (robust standard errors). $^{***}p<0.01$, $^{**}p<0.05$, $^{*}p<0.1$.\par
\end{minipage}
\end{table}

\subsection{\textit{Prompting Fatigue} and Query Specificity}\label{app:attention-specificity}

This section examines how the consequences of \textit{prompting fatigue} depend on whether the participant's queries name a specific target. To this end, we first manually construct a variable that denotes search specificity. To do this, we examine each free-text query submitted during the task and code it as \textit{specific} if it names a concrete entity, that is, an artist, band, performer, team, or a show, musical, play, or film title, and as \textit{broad} if it uses only genres, categories, activity types, moods, or attributes such as date, location, or price; queries naming only a venue are coded broad. For example, \emph{``Hamilton''} and \emph{``Yankees game''} are specific, while \emph{``jazz concert''} and \emph{``something fun this weekend''} are broad. All 911 free-text queries in the logs were coded, and every label was manually verified. A participant is labeled \textit{specific} if at least one of their queries is specific ($36\%$ of the sample) and \textit{exploratory} otherwise; participants who never enter free text are labeled exploratory, since filter selections are category-level by construction.

Table~\ref{tab:attention_specificity} re-estimates the two specifications of Table~\ref{tab:attention_spec} with the specificity indicator in place of treatment. Column (1) shows that the reallocation of choices toward the top of the list is not confined to either group. The probability of choosing from the top ten rises with the \textit{CW share} of the session among exploratory participants ($0.68$ per unit of \textit{CW share}, $p < 0.001$) and, marginally, among specific participants ($0.29$, $p = 0.094$), with the difference between the two slopes itself only marginally significant ($p = 0.088$). Column (2) shows that the valuation consequences of ending in the top ten differ sharply. Among exploratory participants, events chosen beyond the top ten are valued \$$6.7$ higher ($p = 0.001$); among specific participants the difference is small and insignificant (\$$2.1$, $p = 0.48$), although the interaction itself is not significant ($p = 0.21$). This is consistent with \textsection\ref{sec:mech-verif}. A specific query results in the desired event being located near the top of the results list, so choosing within the first page is harmless. But with a broad query, position may be uninformative about value, and choosing one of the first results means missing out on potentially better results further down the list. Thus, \textit{prompting fatigue} is costly when the search is exploratory.  

\begin{table}[htbp]\centering
\caption{\textit{Prompting Fatigue} and Query Specificity}\label{tab:attention_specificity}
\small
\sbox{\regtablebox}{%
\begin{tabular}{l *{2}{>{\centering\arraybackslash}m{3.0cm}}}
\toprule
\multicolumn{1}{r}{Dep.\ var.:} & \textit{Chose from top 10} & \textit{Valuation} \\
\cmidrule(lr){2-2}\cmidrule(lr){3-3}
 & (1) & (2) \\
\midrule
\textit{CW share} of session & 0.68$^{***}$ &  \\
 & (0.15) &  \\
\addlinespace
Chose beyond top 10 &  & 11.47$^{***}$ \\
 &  & (3.58) \\
\addlinespace
Specific queries & 0.23$^{***}$ & 12.84$^{***}$ \\
 & (0.08) & (4.10) \\
\addlinespace
\textit{CW share} $\times$ Specific & $-$0.39$^{*}$ &  \\
 & (0.23) &  \\
\addlinespace
Beyond top 10 $\times$ Specific &  & $-$7.90 \\
 &  & (6.21) \\
\addlinespace
\midrule
Observations & 362 & 362 \\
\midrule
\multicolumn{3}{l}{\textit{Within-group effect (col.\ 1: probability per unit of CW share; col.\ 2: \$, AME):}} \\ \addlinespace
\quad Exploratory queries & 0.68$^{***}$ & 6.69$^{***}$ \\
 & (0.15) & (2.09) \\
\addlinespace
\quad Specific queries & 0.29$^{*}$ & 2.12 \\
 & (0.18) & (2.99) \\
\addlinespace
\quad Pooled & 0.58$^{***}$ & 4.21$^{**}$ \\
 & (0.11) & (1.69) \\
\addlinespace
\bottomrule
\end{tabular}}%
\usebox{\regtablebox}\par
\vspace{3pt}
\begin{minipage}{\wd\regtablebox}
\footnotesize\linespread{0.9}\selectfont
\textit{Notes:} Both specifications mirror Table~\ref{tab:attention_spec}, with the specificity indicator in place of treatment. Column (1): linear probability model of choosing an event from the first ten positions of the results list on the \textit{CW share} of the session, interacted with the specific-queries indicator (HC1 standard errors in parentheses). Column (2): interval regression of valuation (each price-list response enters as its revealed interval; the top category is censored above \$50) on an indicator for choosing beyond the first ten positions, interacted with the specific-queries indicator (robust standard errors); AME denotes the average marginal effect on $E[\text{valuation}]$. $^{*}p<0.1$, $^{**}p<0.05$, $^{***}p<0.01$.\par
\end{minipage}
\end{table}

\end{document}